\documentclass[12pt]{amsart}
\usepackage{amssymb, amsthm, amsfonts, latexsym}
\usepackage[leqno]{amsmath} 
\usepackage{natbib}
\usepackage{mathtools}
\usepackage{enumitem}
\usepackage{needspace}
\usepackage{array}
\usepackage{graphicx,booktabs,tabularx,longtable,pdflscape,float}
\usepackage{xurl}
\usepackage[multiple]{footmisc}
\usepackage{comment}

\usepackage[left=1.25in, top=1.25in, right=1.25in, bottom=1.25in]{geometry}
\usepackage{setspace}
\usepackage{tgpagella}
\usepackage{microtype}

\usepackage{xcolor}
\usepackage{tikz}

\usepackage{etoolbox}
\patchcmd{\section}{\scshape}{\bfseries}{}{}
\makeatletter
\renewcommand{\@secnumfont}{\bfseries}
\makeatother

\theoremstyle{plain}
\newtheorem{theorem}{Theorem}
\newtheorem{corollary}{Corollary}

\newtheorem{proposition}{Proposition}
\newtheorem*{proposition*}{Proposition}

\newtheorem{example}{Example}
\newtheorem{assumption}{Assumption}
\theoremstyle{definition}
\newtheorem{definition}{Definition}

\newtheoremstyle{boldremark}
 {0.5\topsep}{0.5\topsep}
 {\normalfont}{}
 {\bfseries}{.}
 {.5em}{}
\theoremstyle{boldremark}

\usepackage[
 colorlinks=true,
 linkcolor=blue,
 citecolor=blue,
 urlcolor=blue,
 filecolor=blue,
 hyperfootnotes=false,
 hyperindex,
 breaklinks
]{hyperref}

\newcommand{\Ch}{C^{h}}

\usepackage{url}
\begin{document}
\raggedbottom
\author[Aygün]{Orhan Aygün$^\textrm{1}$}\thanks{$^\textrm{1}$ Boğaziçi University}

\author[Turhan]{Bertan Turhan$^\textrm{2}$}\thanks{$^\textrm{2}$ Iowa State University}

\date{\today}

\title{Affirmative Action in India with Hierarchical Reservations}
\thanks{\textit{First version: August, 2019}. We presented some results of this paper in an earlier working paper, ``\textit{Affirmative action in India: Restricted strategy space, complex constraints, and direct mechanism design}." One of the central notions of this paper, the hierarchical choice rule, was first introduced in the third chapter of \cite{aygun2014three} and in \cite{aygun2017verimli}. \newline  We are especially grateful to Oguzhan Celebi, Manshu Khanna and Bumin Yenmez for their extensive comments and suggestions. We also thank Laura Doval, Faruk Gul, Ran Shorrer, and the participants of the 2026 SAET Conferences (Rio de Janeiro, Brasil) for their feedback. 
}

\begin{abstract}
\onehalfspacing
We study the design of affirmative action mechanisms focusing on India's system of horizontal reservations within vertical reservations. We consider a \textit{one-to-all policy}, under which an applicant eligible for multiple horizontal categories simultaneously contributes toward each category's minimum guaranty. When horizontal reservations have a hierarchical structure, we introduce a choice rule that selects the unique merit-optimal set of applicants subject to these guaranties, satisfies the substitutes condition and size monotonicity. In a matching-with-contracts framework, we show that the generalized deferred acceptance mechanism is stable, strategy-proof, and eliminates justified envy under aggregate choice rules using the hierarchical choice rule in each vertical category. We establish a fundamental trade-off between merit and affirmative action between the one-to-all and one-to-one counting policies.  
\end{abstract}

\begin{titlepage}
    \maketitle
\end{titlepage}

\section{Introduction}

India has implemented the world's most complex affirmative action program to allocate government jobs and university seats since the 1950s. The Indian affirmative action program has been implemented through a highly regulated reservation system that consists of  \textit{vertical} and \textit{horizontal} reservations.\footnote{These terms were first used in the Supreme Court of India (SCI)'s judgment in Indra Sawhney vs. Union of India (1992), which is available at https://indiankanoon.org/doc/1363234/.} The Supreme Court of India (SCI) has been regulating the implementation of both types of reservation through successive verdicts since 1992. Reservations for Scheduled Castes (SC), Scheduled Tribes (ST), Other Backward Classes (OBC) and Economically Weaker Sections (EWS) are called \textbf{vertical} reservations. These groups reserve 15\%, 7.5\%, 27\%, and 10\% of positions, respectively. Individuals not belonging to these categories are referred to as members of the \textbf{general category}.\footnote{In 2019, the Indian legislative body instituted a 10\% vertical reservation for a subset of the general category, the Economically Weaker Section (EWS), defined by an annual income threshold below Rs. 8 lakhs.} The residual 40.5\% of available positions falls into the \textbf{open category}. 

Mandated by the SCI in the Indra Sawhney (1992), vertical reservations have operated as \textbf{over-and-above}. That is, open category positions secured by individuals belonging to SC, ST, OBC, or EWS are not deducted from their respective vertical reservations. To implement vertical reservations as over-and-above, institutions fill open category positions, considering all applicants, before filling vertically reserved positions.

Horizontal reservations are designated for other marginalized groups, such as women and individuals with disabilities. The judicial mandates require that horizontal reservations be implemented in a compartmentalized fashion within each vertical category, including the open category.\footnote{\cite{sonmez2022affirmative} discuss the concurrent implementation of vertical and horizontal reservations from a market design perspective.} The SCI has also mandated that horizontally reserved slots should be allocated before any unreserved positions, serving as \textbf{minimum guarantees}.

In Saurav Yadav v. State of Uttar Pradesh (2020), the Court required horizontal reservations to be implemented in the open category by considering all applicants.\footnote{\cite{sonmez2022informed} provides a detailed analysis of this judgment.} When a candidate belongs to both vertical and horizontal reserved categories and scores high enough to qualify for an open category position, she must be considered for horizontally reserved positions in the open category. This judgment removes the ambiguity on the implementation of horizontal reservations in the open category. 

 However, the implementation of horizontal reservations within vertical categories remains ambiguous when applicants possess \textbf{multiple horizontal types}. The law is completely silent on whether an applicant who qualifies for multiple horizontal reservation categories is counted against all of them or only one. In practice, both one-to-all and one-to-one approaches have been used. The law leaves this decision to the implementing authority. For example, the state of Odisha mandates one-to-all implementation of horizontal reservations at the state level. 
 
Unlike \cite{sonmez2022affirmative}, who study the problem in a \textbf{one-to-one} counting scheme, whereby each candidate is counted toward at most one horizontal trait, we adopt a \textbf{one-to-all} counting scheme, under which a candidate may simultaneously satisfy the requirements of multiple horizontal traits. The one-to-one policy may achieve a strictly higher level of affirmative action, albeit potentially at the expense of merit. The choice between these frameworks therefore reflects a fundamental trade-off between merit and affirmative action: the one-to-all policy suits environments where merit is the primary objective, whereas the one-to-one policy is more appropriate when greater representation of horizontally reserved groups takes precedence.

 Under the one-to-all matching policy, complementarities between applicants may arise, thus obviating the existence of a stable matching. However, when horizontal reservations adhere to a hierarchical structure (also known as "laminar"), stable allocations can be achieved.\footnote{See also \cite{huang2010classified}, \cite{fleiner2012matroid}, and \cite{gonczarowski2020matching}.} The Rights of Persons with Disabilities (PwD) Act (2016) in India mandates a 4\% total reservation for PwD in government employment and educational institutions.\footnote{The Act is available at https://niepvd.nic.in/the-rights-of-persons-with-disabilities-rpwd-act-2016/.} Within this aggregate quota, 1\% is reserved for individuals with blindness or low vision and 1\% for those who are deaf or hard of hearing. Moreover, Tamil Nadu's Public Service Commission (TNPSC) system reserves seats for \textit{Poor Widows} within the \textit{Women} reservation. This is another application of hierarchical horizontal reservations.\footnote{The official notification of Tamil Nadu Public Service Commission can be found at \url{https://web.archive.org/web/20260221211454/https://tnpsc.gov.in/Document/english/GRP2_11_2025_ENGLISH.pdf} (last accessed on March 23, 2026).}  The theoretical framework developed in this paper applies directly to both of these structures.

Hierarchical reservations are common throughout the world. The Brazil 2012 Quota Law established a three-tier hierarchical reserve system for federal university admissions at the program level.\footnote{See \cite{aygun2021college} for a detailed analysis of Brazil's 2012 Quota Law and its implementation. Their analysis is based on slot-specific priorities of \cite{kominers2016matching}. See also \cite{avataneo2021slot} for generalization of slot-specific priorities with transfers.} The first tier reserves 50\% of seats for students who completed their entire secondary education in public schools, the remaining 50\% are allocated through open competition. Within the public school quota, seats are subdivided by family income. Half are reserved for students with per capita family income below 1.5 times the minimum wage, while the other half are reserved for the rest. Within each income category, seats are further allocated by race and ethnicity, with the proportion of seats reserved for self-declared Black, Brown, and Indigenous, matching their representation.

 We first consider a single vertical category and formulate the \textbf{hierarchical choice rule} that minimally deviates from fully meritorious outcomes under the one-to-all counting scheme. Theorem \ref{thm:1}  shows that the hierarchical choice rule is not dominated by merit among the choice rules that satisfy horizontal reservations. The proposition \ref{prop:properties} states that the hierarchical choice rule satisfies the substituent property and the size monotonicity. 

 Next, for a given institution, we introduce an \textbf{aggregate choice rule} in which applicants are selected within a vertical category via a hierarchical choice rule. In Theorem \ref{thm:aggchoice}, we show that the aggregate choice rule is fair. Finally, for the centralized matching market that consists of multiple institutions, we show in Theorem \ref{thm:mechanism} that the generalized deferred acceptance (DA) mechanism is the unique stable mechanism that is strategy-proof and eliminates justified envy.\footnote{In Proposition \ref{prop:independence}, we show that stability with respect to fair choice rules and eliminating justified envy are independent properties.} 

We extend our model to formulate OBC de-reservations that are implemented in admissions to public universities. We introduce a new \textit{aggregate choice rule with transfers} that uses the hierarchical choice rule in each vertical category and that allows unfilled OBC positions to be filled at the end by considering all remaining applicants. We show in Theorem \ref{extensionresult1} that this new choice rule is fair. Moreover, in Theorem \ref{extensionresult2}, we show that the generalized DA mechanism under the new aggregate choice rule is stable, strategy-proof, and eliminates justified envy.

In section 6, we make a precise comparison between one-to-all and one-to-one counting schemes. We show that every set of applicants that is feasible under one-to-one counting is also feasible under one-to-all counting (Proposition \ref{prop:onevsall}), so the hierarchical choice rule (weakly) merit-dominates every choice rule that satisfies horizontal reservations under the one-to-one policy (Corollary~\ref{cor:onetoone}) as well. Proposition~\ref{prop:degree} then quantifies the difference. One-to-one counting always admits at least as many beneficiaries of horizontal reservations as the sum of the horizontal quotas, and when one reservation is nested within another, it requires strictly more beneficiaries than one-to-all counting does. The one-to-one policy therefore delivers a greater degree of affirmative action and the one-to-all policy delivers greater fidelity to merit. Finally, Section~\ref{sec:nonhier} shows that the hierarchical structure is essential. When horizontal reservations overlap without being nested, no choice rule that satisfies horizontal reservations and does not sacrifice merit unnecessarily satisfies the substitutes property, and there are markets with two institutions in which no stable matching exists (Proposition~\ref{prop:nonhier}).

Unlike previous work in resource allocation problems in India, which assumes indifference between open and reserved positions among beneficiaries, we model preferences over a richer domain that allows individuals to express preferences over pairs of institutions and vertical position categories. Preferences over position types are heterogeneous and reflect competing social and economic concerns. Some applicants eligible for reserved-category positions may prefer open category admission because of the perceived stigma associated with affirmative action. Empirical evidence suggests that such concerns can influence participation in reservation programs. In particular, \cite{gille2013stigma} finds that, conditional on household wealth, members of OBC subcastes with higher social standing within their villages are less likely to apply for educational reservations, a pattern consistent with reservation-related stigma. Moreover, \cite{siddique2011evidence} documents caste-based discrimination in India's private-sector labor market, while qualitative evidence suggests that reservation beneficiaries may face negative perceptions in educational and professional environments.\footnote{See, for example, \url{https://restofworld.org/2022/tech-india-caste-divides/}.} These findings suggest that some applicants may associate reserved-category admission with reputational or psychological costs and consequently prefer an open category position at the same institution. Importantly, this preference need not reflect an actual labor-market premium for open category admission. Perceived stigma alone may be sufficient to generate it.
 
In contrast, some applicants from reserved categories may prefer reserved seats to open category positions. This preference is largely rooted in the financial benefits that come with a reserved-category position. Federal and state policies frequently provide scholarships, fee waivers, and other forms of financial assistance exclusively to holders of reserved seats. Importantly, access to open category positions also correlates with socioeconomic status, since financially advantaged applicants disproportionately invest in private coaching and test-preparation services. As a result, having a reserved seat conveys information about the socioeconomic background of an applicant. The associated financial support may outweigh any reputational costs for economically constrained individuals.

The remainder of the paper is organized as follows. Section 2 presents the notation for a single vertical category, introduces the hierarchical choice rule, and establishes its main properties under hierarchical horizontal reservations. Section 3 presents the model for a given institution in the context of India's affirmative action system and studies its aggregate choice rule. Section 4 introduces the centralized matching market that consists of multiple institutions and studies the generalized deferred acceptance mechanism. Section 5 extends the framework to accommodate soft reserves. Section 6 compares the one-to-one and one-to-all horizontal counting policies. Section 7 examines the role of hierarchical reservations and the implications of relaxing the hierarchical structure. Section 8 situates our contributions within the related literature, and Section 9 concludes. All proofs are relegated to the Appendix.

\section{A Single Vertical Category}

Let $I=\{i_{1},...,i_{n}\}$ be a finite set of individuals. We first consider a single vertical category $v$ with $q^{v}$ positions. Individuals are ranked according to a strict merit ordering $\succ$. Throughout this section, we take as given the set of individuals who are eligible for category $v$. Eligibility for vertical categories is specified in Section 3.

\subsection{Horizontal Reservations}

Let $H=\{h_{1},\ldots,h_{L}\}$ be the set of horizontal types. The correspondence $\rho:\:I\rightrightarrows H\cup\{\emptyset\}$ represents individuals' horizontal type eligibility. $\rho(i)=\emptyset$ means that $i$ has no horizontal type. Given a set $A \subseteq I$, we write $A_h \equiv A\cap \rho^{-1}(h)$ for the individuals in $A$ who have type $h$ and $A_H \equiv \bigcup_{h\in H} A_h$ for those who have at least one horizontal type.

We denote by $\kappa_{v}^{j} \in \mathbb{Z}_{+}$ the number of positions reserved for horizontal type $h_{j}\in H$ at vertical category $v$.

We say that horizontal reservations are \textbf{hierarchical}, if for any pair of types $h,h^{'}\in H$ such that $\rho^{-1}(h)\cap\rho^{-1}(h^{'})\neq\emptyset$, either $\rho^{-1}(h)\subset\rho^{-1}(h^{'})$ or $\rho^{-1}(h^{'})\subset\rho^{-1}(h)$. When  $\rho^{-1}(h^{'})\subset\rho^{-1}(h)$, we say that the horizontal type $h$ \textit{contains} $h^{'}$.

\subsection{Choice Rules and Their Properties}

A category-$v$ choice rule $C^{v}: 2^{I} \longrightarrow2^{I}$ describes how category $v$ would choose from any set of individuals. For all $A\subseteq I$, the choice rule $C^{v}$ is such that $C^{v}(A)\subseteq A$ and $\mid C^{v}(A)\mid \leq q^{v}$.

A choice rule $C^{v}: 2^{I} \longrightarrow2^{I}$ satisfies the \textit{substitutes} property if, for all $i,j\in A \subseteq I$ and $B\subseteq A \setminus \{i,j\}$,
	\[
	j\notin C^{v}(B\cup\{j\})\;\Longrightarrow\;j\notin C^{v}(B\cup\{i,j\}).
	\]

 A choice rule $C^{v}: 2^{I} \longrightarrow2^{I}$ satisfies \emph{size monotonicity} if, for all individuals $i\in A$ and sets of individuals $B\subseteq A$, we have \[
	\mid C^{v}(B)\mid\leq\mid C^{v}(B \cup \{i\}) \mid.
	\]

The substitutes property and size monotonicity jointly imply the \textit{irrelevance of rejected contracts} (IRC) condition (\cite{aygun2013matching}). A choice rule $C^{v}$ satisfies the IRC condition if, for any $B\subseteq I$ and any $i\in I \setminus B$,  
$$
i\notin C^{v}(B\cup \{i\}) \text{ implies } C^{v}(B)=C^{v}(B\cup \{i\}).
$$

\subsection{Comparison based on merit}

Our main motivation for adopting the one-to-all horizontal matching framework is the ability to design a choice rule that admits the most meritorious set of applicants in each vertical reserve category. We first define a criterion, a version of the domination criterion defined in \cite{gale1968optimal}, to compare two sets of individuals based on merit scores.

\begin{definition}\label{meritdomination}
	Fix a strict merit ordering $\succ$ of applicants, and let $A,B\subseteq I$ with $|A|\geq |B|$. We say that $A$ \emph{weakly merit-dominates} $B$, denoted $A\succeq_{M} B$, if there exists an injection $g:B\to A$ such that $g(b)\succeq b$ for every $b\in B$. We say that $A$ \emph{merit-dominates} $B$, denoted $A\succ_{M} B$, if $A\succeq_{M} B$ and either $|A|>|B|$ or $g(b)\succ b$ for some $b\in B$ for an injection witnessing weak merit domination.
    \end{definition}

Equivalently, write $A=\{a_1,\ldots,a_m\}$ and $B=\{b_1,\ldots,b_n\}$ in decreasing order of merit, with $m\geq n$. Then $A\succeq_{M} B$ if and only if $a_k\succeq b_k$ for $k=1,\ldots,n$. Merit domination is obtained if, in addition, $m>n$ or $a_k\succ b_k$ for some $k\leq n$. For equal-cardinality sets, merit domination is, therefore, exactly the usual coordinatewise merit improvement with at least one strict inequality.

Next, we extend the merit-based comparison criterion defined above to compare acceptant choice rules. A choice rule $C$ is \textit{acceptant} if, for a given capacity $x\in \mathbb{Z}_{+}$ and for each set of individuals $A \subseteq I$, $|C(A)|=min\{|A|,x\}$.

\begin{definition}\label{meritcomparisonchoicerules}
Let $\widetilde C$ and $C$ be acceptant choice rules with the same capacity. We say that $\widetilde C$ \emph{weakly merit-dominates} $C$, denoted $\widetilde C\succeq_{M} C$, if $\widetilde C(A)\succeq_{M} C(A)$ for every set of individuals $A\subseteq I$. We say that $\widetilde C$ \emph{merit-dominates} $C$, denoted $\widetilde C\succ_{M} C$, if $\widetilde C\succeq_{M} C$ and $\widetilde C(A)\succ_{M} C(A)$ for at least one set of individuals $A\subseteq I$.
\end{definition}

Note that this domination relation does not induce a complete binary relation.\footnote{Consider the example where $q=2$ and $I=\{i_{1},i_{2},i_{3},i_{4}\}$. Suppose the individuals' test scores are ordered from highest to lowest as $i_1, i_2, i_3, i_4$. Let $\widetilde{C}$ and $C$ be q-responsive choice rules such that $\widetilde{C}(I)=\{i_{1},i_{4}\}$ and $C(I)=\{i_{2},i_{3}\}$. According to our domination criterion, the sets $\{i_{1},i_{4}\}$ and $\{i_{2},i_{3}\}$ do not dominate each other.}

Since India mandates implementing horizontal reservations in each vertical category, we are interested in choice rules that satisfy horizontal reservations. 

\begin{definition}\label{def:feas}
Let $A\subseteq I$ be a set of individuals who are eligible for category $v$. 
A category-v choice rule $C^{v}$ \emph{satisfies horizontal reservations} if $C^{v}A)\in \mathcal{F}(A)$ for every $A$ with $\mathcal{F}(A)\neq\emptyset$, where
the family of \emph{HR-feasible sets} is defined as
\[
\mathcal{F}(A)\;\equiv\;\Bigl\{\,B\subseteq A \;:\; |B|=\min\{q,|A|\}\ \text{ and }\ |B\cap A_h|\ \ge\ \min\{\kappa_h,|A_h|\}\ \text{ for every } h\in H \Bigr\}.
\]

\end{definition}

Note that the requirement $|B\cap A_h|\ge \min\{\kappa_h,|A_h|\}$ is the \textbf{one-to-all count}. An individual with several horizontal types is counted towards each of them.

\begin{definition}\label{meritundominatedchoice}
A category-v choice rule $C^{v}$ is \textbf{merit-undominated} among the choice rules that satisfy horizontal reservations if, for every set of applicants $A\subseteq I$, $C^{v}(A)$ is not merit-dominated by any set $B \subset A$ that satisfies horizontal reservations. 
\end{definition}

\subsection{Hierarchical Choice Rule}

We restrict attention to hierarchical horizontal reservations within the one-to-all horizontal type matching framework. We introduce the hierarchical choice rule, which operates sequentially. It first admits applicants who qualify for minimal (innermost) types first and then iteratively continues in order of decreasing qualification scope. 

We write $h'\prec h$ when $\rho^{-1}(h')\subsetneq \rho^{-1}(h)$. Let $H^{\max}$ be the set of maximal types, and let $\operatorname{ch}(h)$ be the set of maximal proper subtypes of $h$. We maintain the following feasibility requirement for hierarchical reservations throughout.

\begin{assumption}\label{ass:feasible}
The horizontal quotas satisfy
\[
\sum_{h\in H^{\max}}\kappa_h\le q^{v}
\text{  and  }
\sum_{h'\in\operatorname{ch}(h)}\kappa_{h'}\le\kappa_h
\quad\text{for every nonminimal }h\in H.
\]
\end{assumption}

Assumption~\ref{ass:feasible} ensures that horizontal-reservation selections never exceed total capacity. At most $\kappa_h$ applicants are selected while processing the sub-tree rooted at any $h$. Since distinct maximal types have disjoint eligibility sets, at most $\sum_{h\in H^{\max}}\kappa_h\le q^{v}$ applicants are selected before the final merit-based step. The assumption does not require that every quota be filled when eligible applicants are scarce.

For an applicant set $A\subseteq I$, let $A_h=A\cap \rho^{-1}(h)$. Let $(H_1,\ldots,H_K)$ be the layers obtained by repeatedly removing the minimal types under $\prec$. $H_1$ consists of the minimal types and $H_k$ consists of the minimal types remaining after layers $1,\ldots,k-1$ are removed. For $B\subseteq A$ and $m\ge0$, let $\operatorname{Top}_m(B)$ be the $\min\{m,|B|\}$ highest-merit applicants in $B$.

\paragraph{\textbf{Hierarchical choice rule}}
Set $S_0=\varnothing$. For each $k=1,\ldots,K$ and $h\in H_k$, define the residual horizontal quota and the new selections by
\[
r_h(S_{k-1})=\max\{\kappa_h-|S_{k-1}\cap A_h|,0\},
\qquad
B_h^k=\operatorname{Top}_{r_h(S_{k-1})}\bigl(A_h\setminus S_{k-1}\bigr),
\]
Update
\[
S_k=S_{k-1}\cup\bigcup_{h\in H_k}B_h^k.
\]
Finally, fill any remaining positions by merit:
\[
 C^h(A;q^{v},\kappa,\rho)
=S_K\cup\operatorname{Top}_{q^{v}-|S_K|}(A\setminus S_K).
\]
Applicants selected at any stage count toward every horizontal type for which they are eligible. The selection sets for types in the same layer are disjoint, and Assumption~\ref{ass:feasible} guarantees $|S_k|\le q^{v}$ at every stage. If $|S_k|=q^{v}$, all subsequent selections are empty.

We characterize the hierarchical choice rule as the best rule with respect to merit among those that satisfy horizontal reservations. 

\begin{theorem}\label{thm:1}
For every $A\subseteq I$ eligible for category $v$ and every $B\in\mathcal{F}(A)$ with $B\neq C^{h}(A,\kappa,\rho)$, the set $C^{h}(A,\kappa,\rho)$ merit-dominates $B$. In particular, $C^{h}$ is the unique choice rule that is merit-undominated among the choice rules that satisfy horizontal reservations.
\end{theorem}

The hierarchical choice rule $C^{h}$ satisfies important properties that are deemed standard in the market design literature.

\begin{proposition}\label{prop:properties}
    $C^{h}$ satisfies the substitutes property and size monotonicity. 
\end{proposition}

\section{An Institution and Its Aggregate Choice Rule}

We now consider an institution $s$ that fills positions in several vertical categories. $\overline{q}_{s}$ denotes the number of positions available in the institution $s$. 

We let $R=\{SC,ST,OBC, EWS\}$ denote the set of vertical reservation categories. Institution $s$ sets aside a certain number of positions for each of the vertical reserve categories. Let $q_{s}^{SC}$, $q_{s}^{ST}$, $q_{s}^{OBC}$, and $q_{s}^{EWS}$ be the number of positions reserved for SC, ST, OBC, and EWS, respectively. We assume that $\overline{q}_s \geq \underset{r\in R}{\sum} q_{s}^r$. The residual $q_{s}^{o}=\overline{q}_{s}-\underset{r \in R}{\sum}q_{s}^{r}$ positions are called the \textit{open category} positions, where $o$ denotes the open category. We denote the set of position types $V=\{o,SC,ST,OBC,EWS\}$. We denote the vector of the capacities of vertical categories at $s$ by $q_{s}=(q_{s}^{o},q_{s}^{SC},q_{s}^{ST},q_{s}^{OBC},q_{s}^{EWS})$.

The function $t:I \rightarrow R \cup \{ g\}$ denotes the membership of the vertical category of individuals, where $g$ denotes \textit{the general category}. For each $i\in I$, $t(i)$ denotes the category individual $i$ belongs to. We say $t(i)=g$ only if $t(i) \notin R$. We denote a vertical category membership profile by $T=\left(t_{i}\right)_{i \in I}$, and let $\mathcal{T}$ be the set of all possible profiles. 

Individuals belonging to the general category (g) are eligible only for open category (o) positions. In contrast, individuals belonging to a vertically reserved category are eligible for both open category positions and positions reserved for their respective categories. Thus, open category positions are available to all individuals, regardless of their vertical category membership, whereas positions reserved for a particular vertical category are available exclusively to members of that category.

Should reserved-category candidates abstain from declaring their membership, they are considered exclusively for the open category. Failure to declare membership in any vertical category defaults candidates to the general category. The legal framework in India requires the disclosure of vertical category membership to be discretionary. 

Institution $s$ has a strict ranking of individuals $\succ_{s}$. In India, $\succ_{s}$ is induced by the ranking of the test-score. Given the master ranking $\succ_{s}$, for each vertical category $r\in R$, the strict ranking of individuals for category-$r$ positions is obtained simply as follows: 

\begin{itemize}
    \item $i \succ_{s}^{r} j$ if and only if $i \succ_{s} j$, for all $i,j \in I$ such that $t(i)=t(j)=r$, and 
    \item $i \succ_{s}^{r} \emptyset \succ_{s}^{r} j$, for all $i,j \in I$ such that $t(i)=r$ and $t(j) \neq r$.
\end{itemize}

Note that $\emptyset \succ_{s}^{r} j$ means that individual $j$ is not eligible for category-$r$ positions. 

The vector $\kappa_{v}^{s}\equiv(\kappa_{v}^{1},...,\kappa_{v}^{L}) \in \mathbb{Z}_{+}^{L}$ denotes the horizontal reservations in vertical category $v\in V$ at $s$. Let $\kappa_{s} \equiv\left\{ \kappa_{v}^{s}\right\} _{v\in V} \in \mathbb{Z}_{+}^{|V| \times L}$ denote the horizontal reservations in $s$.\footnote{We assume that individuals do not have preferences over horizontal types, since there is no evidence that they care about the horizontal types under which they are admitted. All discussions of reservation policy in India focus on vertical reservations. Perhaps more importantly, the matching outcomes do not specify the horizontal categories under which individuals are assigned. }

\subsection{Contracts} 

There is a finite set of contracts between individuals and institution $s$, denoted $X_{s}$. Taking into account eligibility requirements, we define the set of all contracts of institution $s$ as 

\begin{equation*}
X_{s} = \left( \underset{t(i)=g}{\bigcup}\{i\} \times \{s\} \times \{o\}\right) \bigcup \left(\underset{t(i) \in R}{\bigcup}\{i\} \times \{s\} \times\{r,o\}\right).
\end{equation*}

Each contract $x \in X_{s}$ is between an individual $\mathbf{i}(x)$ and institution $s$ and specifies a vertical category $\mathbf{t}(x)$ in which the individual
$\mathbf{i}(x)$ is admitted. 

For any set of contracts $Y\subseteq X_{s}$, let $\mathbf{i}(Y)\equiv \bigcup_{x\in Y}\{\mathbf{i}(x)\}$ be the set of individuals who have a contract associated with them in $Y$. We also define $Y_{i}\equiv \{x\in Y\mid i=\mathbf{i}(x)\}$  for each $i\in I$. 

\subsection{Aggregate Choice Rules and Their Properties}

Institution $s$ has multiunit demand and is endowed with an \textbf{aggregate choice rule} $C_{s}: 2^{X_s} \longrightarrow2^{X_{s}}$ that describes how $s$ would choose from any set of contracts. For all $Y\subseteq\ X_{s}$, the choice rule $C_{s}$ is such that $C_{s}(Y)\subseteq Y$ and $\mid C_{s}(Y)\mid \leq q_{s}$. The substitutes property, size monotonicity, and the IRC condition are defined for $C_{s}$ analogously, with sets of contracts in $X_{s}$ in place of sets of individuals.

The fairness of a choice rule is an important design criterion (\cite{aygun2021college}). 

\begin{definition}\label{fairchoice}
	An aggregate choice rule $C_{s}$ is $\mathbf{fair}$ if, for any set of contracts $Y\subseteq X_{s}$ and any $x\in Y$, the chosen set $C_{s}(Y)$ satisfies the following: If $\mathbf{i}(x)\notin \mathbf{i}[C_{s}(Y)]$, then for every $y\in C_{s}(Y)$ one of the following holds:
\begin{enumerate}
        \item $\mathbf{i}(y)\succ_{s}\mathbf{i}(x)$, 
        \item $t(x)\neq t(y)$, 
        \item $\rho(\mathbf{i}(y))\setminus\rho\left(\mathbf{i}(x)\right)\neq\emptyset$. 
\end{enumerate}

\end{definition}

If an individual is not selected -- that is, none of her contracts are accepted -- then those who are chosen must either possess superior merit scores or be beneficiaries of vertical or horizontal reservations that the individual who was not chosen lacks. 

In the context of Indian resource allocation, the fairness of a choice rule requires that merit scores are given due consideration, subject to legally imposed vertical and horizontal reservations. When comparing two individuals with identical vertical and horizontal types, priority is given to the one with the higher merit score.

\subsection{Aggregate Choice Rule \texorpdfstring{$\mathcal{C}^{h}$}{C	extasciicircum h}}

The aggregate choice rule $\mathcal{C}^{h}$ is defined as 
$\mathcal{C}^{h}\equiv\left(\left(C_{v}^{h}\right)_{v\in V},q_{s}\right)$, where the precedence sequence is $o-SC-ST-OBC-EWS$. According to $\mathcal{C}^{h}$, each vertical category $v\in V$ uses the hierarchical choice rule $C^h$ to fill its positions given the capacity vector for horizontal reservations $\kappa_{s}$ and the horizontal type correspondence of individuals $\rho$.  
	
Our next result states that the aggregate choice rule $\mathcal{C}^{h}$ satisfies the fairness condition defined for choice rules.

\begin{theorem}\label{thm:aggchoice}
	The aggregate choice rule $\mathcal{C}^{h}$ is fair.
\end{theorem}

This property is especially important in decentralized markets in which each institution runs its own choice rule to select applicants independently. 

\section{Centralized Matching Market}

We now consider a centralized matching market that consists of multiple institutions. Let $S=\{s_{1},...,s_{m}\}$ be a finite set of institutions. Each institution $s\in S$ is as described in Section 3, with its own capacities $q_{s}$, strict ranking of individuals $\succ_{s}$, horizontal reservations $\kappa_{s}$, and aggregate choice rule $C_{s}$.

\subsection{Contracts} 

The set of all contracts between individuals and institutions is $X=\bigcup_{s\in S}X_{s}$. That is, 

\begin{equation*}
X = \left( \underset{t(i)=g}{\bigcup}\{i\} \times S \times \{o\}\right) \bigcup \left(\underset{t(i) \in R}{\bigcup}\{i\} \times S \times\{r,o\}\right).
\end{equation*}

Each contract $x \in X$ is between an individual $\mathbf{i}(x)$ and an institution $\mathbf{s}(x)$ and specifies a vertical category $\mathbf{t}(x)$ in which the individual
$\mathbf{i}(x)$ is admitted. 

For any set of contracts $Y\subseteq X$, $\mathbf{i}(Y)$ and $Y_{i}$ are defined as in Section 3. Similarly, we let $\mathbf{s}(Y)\equiv \bigcup_{x\in Y}\{\mathbf{s}(x)\}$. We also define $Y_{s}\equiv \{x\in Y\mid s=\mathbf{s}(x)\}$ for each $s\in S$. For all $Y\subseteq\ X$ and $s\in S$, we write $C_{s}(Y)$ for $C_{s}(Y_{s})$, so that $C_{s}(Y)\subseteq Y_{s}$. 

\subsection{Preferences of Individuals}

The matching outcomes in India specify two pieces of information for each applicant: (i) the institution with which the applicant is matched and (ii) the vertical category under which she/he is admitted. However, Indian authorities solicit applicants' preferences exclusively over institutions, presuming that candidates from reserved categories are indifferent between open- and reserved-category positions within the same institution. However, disclosing vertical-category membership is discretionary and can be leveraged strategically, as the following example illustrates. 

\begin{example}
    Consider an OBC candidate with a rank-ordered list a-b-c over institutions $\{a,b,c\}$. Should she choose to disclose her OBC membership, she would be evaluated in accordance with the following sequence: \[\left(a,Open\right)-\left(a,OBC\right)-\left(b,Open\right)-\left(b,OBC\right)-\left(c,Open\right)- \left(c,OBC\right).\]  
  
In contrast, if she opts to withhold her OBC membership while submitting identical preferences, her evaluation would proceed as follows.  
  
  \[\left(a,Open\right)-\left(b,Open\right)-\left(c,Open\right).\]

If the applicant prefers $(b,Open)$ to $(a,OBC)$, she can choose to hide her category membership. Moreover, some preferences, such as the one below, cannot be expressed according to the current language.
\[\left(a,Open\right)-\left(b,Open\right)-\left(c,Open\right)-\left(a,OBC\right)-\left(b,OBC\right)- \left(c,OBC\right).\]

\end{example}

We model the preferences of individuals over institution and position category pairs. Individuals in the general category have preferences over $S \times \{ o\} \cup\left\{ \emptyset\right\} $. Each member of the category $r$ has a preference over $S \times\left\{ r,o\right\} \cup\left\{ \emptyset\right\} $,
where $\emptyset$ denotes the outside option.

We let $P_{i}$ denote the strict preference relation of individual $i\in I$. We write $(s,v)P_{i}(s^{'},v^{'})$ to mean that individual $i$ strictly prefers admission to institution $s$ through the vertical category $v$ to admission to institution $s^{'}$ through the vertical category $v^{'}$. The \textit{at-least-as-well} relation $R_{i}$ is obtained from $P_{i}$ as follows: $(s,v)R_{i}(s^{'},v^{'})$
if and only if either $(s,v)P_{i}(s^{'},v^{'})$ or $(s,v)=(s^{'},v^{'})$.
A pair $(s,v) \in S \times V$ is \emph{acceptable} to individual $i$ if it is at least as good as the outside option $\emptyset$ and is unacceptable to her if it is worse than $\emptyset$. 

Let $\mathcal{P}_{i}$ denote the set of all possible preferences for individual $i$ and $\mathcal{P}$ denote the set of all preference profiles.

\subsection{Properties of Matchings}

In the context of merit-based object allocation subject to vertical and horizontal reservations in India, we are also interested in matching outcomes that \textbf{eliminate justified envy}. A matching is \textit{feasible} if it respects vertical and horizontal eligibility requirements and institutions' capacities.

\begin{definition}
A feasible matching $Y\subseteq X$ \textbf{eliminates justified envy} if, for any individual j and a contract $x\in Y\setminus Y_j$, $(\mathbf{s}(x),t(x))P_{j} Y_j$ implies $$ \mathbf{i}(x)\succ_{\mathbf{s}(x)}j \text{ or } \rho(j)\nsupseteq\rho(\mathbf{i}(x)).$$ A mechanism $\psi$ \textit{eliminates justified envy}, if for any $P$, $\psi(P)$ eliminates justified envy. 
\end{definition}

A feasible matching eliminates justified envy if an individual envies the assignment of another, then either the former individual has a lower merit score at that institution or the latter individual has a horizontal reservation type that the former does not.\footnote{Note that if $(\mathbf{s}(y),\mathbf{t}(y))P_{\mathbf{i}(x)}(\mathbf{s}(x),\mathbf{t}(x))$ is possible only when the individual $\mathbf{i}(x)$ is eligible for the vertical category $\mathbf{t}(y)$.}
\begin{definition}
    A feasible outcome $Y\subseteq X$ is \textbf{stable} if 
\begin{enumerate}
    \item $Y_{i}R_{i}\emptyset$, for all $i\in I$, 
	\item $C_{s}(Y)=Y_{s}$, for all $s\in S$, and
	\item there is no $Z\subseteq(X \setminus Y)$, such that $Z_{s}\subseteq C_{s}(Y\cup Z)$ for all $s\in\mathbf{s}(Z)$ and $ZP_{i}Y$ for all $i\in\mathbf{i}(Z)$. 
\end{enumerate}
\end{definition}

Stability has been central in matching markets since \cite{gale1962college}. Condition (1) is the standard individual rationality condition for individuals. Condition (2) requires that institutions' selection procedures be respected. This condition ensures compliance with criteria encoded in institutions' choice rules. Condition (3) is the standard \textit{no blocking condition} and requires that there be no set of contracts $Z$ such that all institutions and individuals associated with contracts in $Z$ prefer to receive the contracts in $Z$. 

The following result establishes the independence of two properties of matchings: (i) elimination of justified envy and (ii) stability with respect to fair choice rules.\footnote{\cite{romm2024stability} generalize the definition of justified envy in matching with contracts environments and show that stable allocations might have justified envy. Our definition of fairness for choice rules differs from their definition of strong priority, but the two notions are related.} 

\begin{proposition}\label{prop:independence}
Elimination of justified envy and stability with respect to fair choice rules are independent. 
\end{proposition}

The two notions differ for two reasons. Stability with respect to a fair choice rule enforces the institution's precedence order over position types together with non-wastefulness but treats the position type an individual receives as the institution's decision. A fair rule may assign a high-scoring reserved-category applicant to the open category even when she prefers a reserved position, leaving the reserved position to a lower-scoring applicant, which is justified envy in the sense of Definition~6. Conversely, elimination of justified envy respects individuals' own preferences over position types but says nothing about unfilled positions or the institution's choice rule, so an outcome that leaves a position empty can be envy-free and unstable. The wedge is exactly heterogeneity in preferences over position types, which is the feature of the Indian setting that motivates our preference domain. Theorem~\ref{thm:mechanism} shows that the cumulative offer mechanism under $C^{h}$ nonetheless delivers both properties; the independence concerns the properties, not a trade-off the mechanism faces.

\subsection{Mechanisms and Their Properties}

Given a profile of choice rules $C=(C_{s})_{s\in\ S}$, a \textbf{mechanism} $\psi(\cdot;C)$ maps preference profiles $P=(P_{i})_{i\in\ I}$ to the outcomes. Unless otherwise stated, we assume that institutions' choice rules are fixed. 

A mechanism $\psi$ is \textbf{stable} if $\psi(P; C)$ is a stable outcome for every preference profile $P$. A mechanism $\psi$ \textbf{eliminates justified envy} if $\psi(P; C)$ eliminates justified envy for every preference profile $P$.

A mechanism $\psi$ is \textbf{strategy-proof} if, given a profile of institutions' choice rules $C$, for every $P\in \mathcal{P}$ and for each individual $i\in I$,
there is no $\widetilde{P}_{i} \in \mathcal{P}_{i}$ such that $$\psi(\widetilde{P}_{i},P_{-i};C)P_{i}\psi(P;C).$$

\subsection{Centralized Mechanism}

We propose the generalized deferred acceptance mechanism (also known as the cumulative offer mechanism (COM)) as a direct mechanism, coupled with institutions' aggregate choice rules. Let $\Phi^{h}$ denote the COM under the profile of choice rules $\mathcal{C}^{h}= \left(\mathcal{C}_{s}^{h}\right)_{s\in S}$.

\textbf{Cumulative Offer Mechanism.}
Given the aggregate choice rules of the institutions $\mathcal{C}^h=\left(\mathcal{C}_{s}^{h}\right)_{s\in S}$ and the preferences of individuals $P=(P_i)_{i\in I}$, the outcome of $\Phi^{h}(P, \mathcal{C}^{h})$ is calculated by the cumulative offer process (COP) as follows: 

\paragraph{\textbf{Step 1}}
Some individual $i^{1}\in\ I$ proposes her most-preferred contract $x^{1}\in\ X_{i^{1}}$. Institution $\mathbf{s}\left(x^{1}\right)$ holds $x^{1}$ if $x^{1}\in \mathcal{C}^{h}_{\mathbf{s}\left(x^{1}\right)}\left(\left\{ x^{1}\right\} \right)$,
and rejects $x^{1}$ otherwise. Set $A_{\mathbf{s}\left(x^{1}\right)}^{2}=\left\{ x^{1}\right\} $,
and set $A_{s'}^{2}=\emptyset$ for each $s'\neq\mathbf{s}\left(x^{1}\right)$.

\paragraph{\textbf{Step k}}
Some individual $i^{k}\in\ I$, for whom no institution currently holds a contract, proposes her most-preferred contract that has not yet been rejected $x^{k}\in X_{i^{k}}\setminus \left(\underset{j\in S}{\bigcup} A_j^k\right)$.\footnote{Note that $A^{k}_{j}$ is the cumulative set of contract proposals to institution $j$ up to Step $k$.} Institution $\mathbf{s}\left(x^{k}\right)$ holds the set of contracts in $\mathcal{C}^{h}_{\mathbf{s}\left(x^{k}\right)}\left(A_{\mathbf{s}\left(x^{k}\right)}^{k}\cup\left\{ x^{k}\right\} \right)$ and rejects all other contracts in $A_{s\left(x^{k}\right)}^{k}\cup\left\{ x^{k}\right\} $; institutions $s'\neq\mathbf{s}\left(x^{k}\right)$ continue to hold
all contracts they held at the end of Step $k-$1. Set $A_{\mathbf{s}\left(x^{k}\right)}^{k+1}=A_{\mathbf{s}\left(x^{k}\right)}^{k}\cup\left\{ x^{k}\right\} $,
and set $A_{s'}^{k+1}=A_{s'}^{k}$ for each $s'\neq\mathbf{s}\left(x^{k}\right)$. 

The algorithm ends when no individual can propose. Each institution is assigned the set of contracts that it holds in the last step. The order of contracts offered during the COM does not matter. \cite{hatfield2017restud} shows that the COM outcome is independent of the order of proposals under a rich class of choice rules.

Our next result shows that the generalized DA mechanism under $\mathcal{C}^{h}=(\mathcal{C}^{h}_{s})_{s\in S}$ is the unique mechanism that satisfies the desirable properties. 

\begin{theorem}\label{thm:mechanism}
	$\Phi^{h}$ is the unique rule that is stable with respect to the aggregate choice rule profile $(\mathcal{C}^h_s)_{s\in S}$, strategy-proof for individuals, and eliminates justified envy. 
\end{theorem}

\section{Extension: Institutions' choice rules with soft reserves}

In allocating government jobs and admissions to publicly funded academic institutions, reservations for SC and ST are categorized as \textbf{hard} reserves, meaning that unfilled positions earmarked for these groups are not transferable to other categories. Although reservations for OBC are treated as hard reserves in the context of governmental job allocations, they were suggested and have been implemented as \textbf{soft} reserves in academic admissions, according to the landmark SCI decision in Ashoka Kumar Thakur vs. Union of India (2008).\footnote{The judgment is available at https://indiankanoon.org/doc/1219385/. See \cite{baswanaetal2019} and \cite{aygun2022dereserve} for details of OBC de-reservations.}

\subsection{Aggregate Choice Rule with Transfer}

The aggregate choice rule with transfers $\mathcal{C}^{hT}\equiv\left(\left(C_{v}^{h}\right)_{v\in\mathcal{V}}, C_{D}, q_{s}\right)$ allocates unfilled OBC positions as open category positions at the end, where 
\begin{itemize}
	\item the precedence order is $o-SC-ST-OBC-EWS-D$, where $D$ denotes the "\textit{de-reserved positions}" that are reverted from OBC, and 
	\item $q_{s}=\left(q_{s}^{o},q_{s}^{SC},q_{s}^{ST},q_{s}^{OBC},q_{s}^{EWS},q_{s}^{D}\right)$
	such that $q_{s}^{D}=r_{OBC}$ is the number of vacant OBC positions. 
    \item $C_{D}$ is a responsive choice rule induced by the merit ranking of individuals and chooses the contracts of the remaining highest-scoring individuals. 
\end{itemize}

Note that unfilled OBC positions are allocated only after all other vertically reserved positions have been filled. There are multiple reallocation schemes for vacant OBC positions. \cite{aygun2026IndiaMD} refer to this particular soft reservation scheme as \textit{forward transfer}. \cite{aygun2022dereserve} introduced the \textit{backward transfer} approach, which re-executes the choice rule from the beginning with updated position counts for vertical categories. Forward and backward transfers represent the two extremes of a continuum of soft reservation policies. Our adoption of the forward transfer scheme is not arbitrary. \cite{aygun2022dereserve} demonstrates that forward transfer maximizes merit among all feasible soft reservation schemes. Thus, merit maximization is the objective of both the choice rule within each vertical category and the soft reservation policy.

\begin{theorem}\label{extensionresult1}
The aggregate choice rule $\mathcal{C}^{hT}$ is fair. 
\end{theorem}

Let $\Phi^{hT}$ be the COM under the profile of choice rules $\left(\mathcal{C}_{s}^{hT}\right)_{s\in\mathcal{S}}$.

\begin{theorem}\label{extensionresult2}
 $\Phi^{hT}$ is the unique rule that is stable with respect to the profile of aggregate choice rules $\left(\mathcal{C}_{s}^{hT}\right)_{s\in\mathcal{S}}$, strategy-proof for individuals, and eliminates justified envy. 
\end{theorem}

\section{One-to-all versus one-to-one Horizontal Matching}

The one-to-one counterpart of Definition \ref{def:feas} counts each selected individual toward at most one type. We next formalize the feasibility in one-to-one horizontal matching.

\begin{definition}\label{def:feasone}
$B\subseteq A$ is \emph{one-to-one HR-feasible}, written $B\in\mathcal{F}^{1}(A)$, if $|B|=\min\{q,|A|\}$ and there is an assignment $\sigma:B\to H\cup\{\emptyset\}$ with $\sigma(i)\in\rho(i)\cup\{\emptyset\}$ for every $i\in B$ and $|\sigma^{-1}(h)|\ge \min\{\kappa_h,|A_h|\}$ for every $h\in H$.
\end{definition}

In one-to-one counting, the sets $\sigma^{-1}(h)$ are pairwise disjoint. Thus, unlike one-to-all counting, a candidate eligible for several horizontal types can satisfy the minimum guarantee of at most one type. For some applicant sets the effective one-to-one minimum guarantees may exceed total capacity, in which case $\mathcal{F}^{1}(A)=\varnothing$. Statements comparing the two policies are on populations for which the relevant feasible sets are nonempty.

\begin{proposition}\label{prop:onevsall}
    For every $A$, $\mathcal{F}^{1}(A)\subseteq\mathcal{F}(A)$.
\label{lem:nested}
\end{proposition}

\begin{proof}
If $\sigma$ is such that $B\in\mathcal{F}^{1}(A)$, then for every $h$ we have $\sigma^{-1}(h)\subseteq B\cap A_h$, so $$|B\cap A_h|\ge|\sigma^{-1}(h)|\ge\min\{\kappa_h,|A_h|\}.$$
\end{proof}

 A choice rule $C$ \emph{satisfies horizontal reservations under the one-to-one policy} if $C(A)\in\mathcal{F}^{1}(A)$ whenever $\mathcal{F}^{1}(A)\ne\emptyset$.

\begin{corollary}\label{cor:onetoone}
Under Assumption 1, for every set $A$ and every $B\in\mathcal{F}^{1}(A)$, $\Ch(A;q_v,\kappa,\rho)\succeq_{M} B$, with strict merit domination whenever $B\neq\Ch(A;q_v,\kappa,\rho)$. Consequently, on every input at which a choice rule that satisfies the horizontal reservations under one-to-one policy selects, the hierarchical choice rule weakly merit-dominates that selection. If both rules are defined on a common domain of such inputs and differ on at least one input, the hierarchical choice rule merit-dominates the other.
\end{corollary}

Note that, by Proposition 3, $B\in\mathcal{F}^{1}(A)$ implies $B\in F(A)$. Corollary \ref{cor:onetoone} follows from Theorem 1.

The next result makes precise the sense in which the one-to-one policy generates a greater degree of affirmative action, and shows that the two counting policies differ even in the simplest hierarchical structure, a reservation nested within another.

\begin{proposition}\label{prop:degree}
\emph{(i)} For every $A$ and every $B\in\mathcal{F}^{1}(A)$,
\[
|B\cap A_H|\ \ge\ \sum_{h\in H}\min\{\kappa_h,|A_h|\}.
\]
\emph{(ii)} Suppose $H=\{h,h'\}$ with $\rho^{-1}(h')\subsetneq\rho^{-1}(h)$. Let $\kappa_h,\kappa_{h'}\geq1$, and suppose
\[
q\geq\kappa_h+\kappa_{h'},\qquad
|A_{h'}|\geq\kappa_{h'},\qquad
|A_h|\geq\kappa_h+\kappa_{h'},\qquad
|A\setminus A_h|\geq q-\max\{\kappa_h,\kappa_{h'}\}.
\]
Then both feasible-set correspondences are nonempty and
\[
\min_{B\in F(A)}|B\cap A_h|=\max\{\kappa_h,\kappa_{h'}\},
\qquad
\min_{B\in\mathcal{F}^{1}(A)}|B\cap A_h|=\kappa_h+\kappa_{h'}.
\]
In particular, $\mathcal{F}^{1}(A)\subsetneq F(A)$, and the minimum number of distinct $h$-eligible beneficiaries required by one-to-one counting strictly exceeds the minimum required by one-to-all counting.
\end{proposition}

Note that the comparison in Proposition~\ref{prop:degree} (ii) concerns the \emph{minimum number of distinct beneficiaries required} by the policies. It does not imply that every outcome under one-to-one counting contains more beneficiaries than every outcome under one-to-all counting. 

Proposition~\ref{prop:degree} (ii) is the Rights of Persons with Disabilities Act (2016) in miniature. With $100$ positions, a $4$ percent reservation for persons with disabilities and a $1$ percent reservation for blindness or low vision nested within it, one-to-all counting requires four beneficiaries of whom at least one is blind or has low vision, whereas one-to-one counting requires five. \textit{The statute describes the $1$ percent as lying within the $4$ percent, which is a one-to-all statement}. One-to-one counting silently converts the nested sub-quota into an additional one. Combining Corollary~\ref{cor:onetoone} with Proposition~\ref{prop:degree}, the hierarchical choice rule selects a set that (weakly) merit-dominates every set admissible under one-to-one counting (merit dominates whenever the sets differ), and the one-to-one policy achieves its greater degree of affirmative action at the expense of merit.

\section{Non-hierarchical reservations}\label{sec:nonhier}

We next formalize that if the horizontal reservations are not hierarchical, then one-to-all horizontal matching can generate complementarities between applicants. 

\begin{proposition}\label{prop:nonhier}

Suppose that there exist horizontal types $h_1,h_2\in H$ such that each of $\rho^{-1}(h_1)\cap\rho^{-1}(h_2)$, $\rho^{-1}(h_1)\setminus\rho^{-1}(h_2)$, and $\rho^{-1}(h_2)\setminus\rho^{-1}(h_1)$ is nonempty, and there is an applicant having neither type. Then there exist a capacity, a vector of horizontal quotas, a merit ordering, and an applicant population for which (i) no choice rule that satisfies horizontal reservations and is merit-undominated at every population satisfies substitutability, and (ii) the resulting choice requirements can be embedded in a two-institution matching market with no stable matching.
\end{proposition}

Note that overlapping non-hierarchical eligibility does not imply failure of substitutability or nonexistence of stable matchings for every quota vector or every applicant population. In the proof, an applicant eligible for both $h_1$ and $h_2$ simultaneously meets two minimum guarantees, freeing a position for an applicant of neither type. Removing the multiply eligible applicant eliminates this possibility and forces the institution to use both positions to meet the separate guarantees. The resulting complementarity is the source of the substitutes violation.

\section{Relation to the Literature}

The closest paper to ours is \cite{gonczarowski2020matching}, where the authors define a choice rule similar to our hierarchical choice rule and also prove the substitute property and the size monotonicity. However, the settings are different. In our case, there is no upper quota on horizontal reservations, while \cite{gonczarowski2020matching} has a laminar structure on upper quotas and a stricter structure on lower quotas. 

This paper contributes to the growing literature on the complexities of resource allocation problems in India. \cite{echenique2015control} is the first paper to present Indian affirmative action as an example of controlled school choice. \cite{aygun2017large} explores the specific challenges faced in admissions to the IITs. \cite{aygun2020dynamic} studied the concurrent implementation of vertical reservations and de-reservations in IIT admissions but did not address horizontal reservations. \cite{sonmez2022affirmative} examines affirmative action policies in India through court decisions from a market design perspective. Our paper diverges from theirs in two crucial dimensions. First, we extend the analysis by accounting for individuals' preferences over both institutions and position categories. Moreover, \cite{sonmez2022affirmative} does not incorporate OBC de-reservations into its model, limiting its analysis to government job allocations where OBC de-reservations are not applicable. \cite{aygun2022dereserve} examines the concurrent implementation of vertical reservations and OBC de-reservations, focusing on the recently revamped admissions procedures for technical universities in India. \cite{aygun2022dereserve} introduces backward-transfer choice rules and proposes the Deferred Acceptance (DA) mechanism under these rules.\footnote{See also \cite{aygun2023priority} for a simpler choice rule to implement backward transfers.}

This paper also contributes to the literature on market design studying affirmative action and diversity. The concept of reserves was introduced in the seminal work of \cite{hafalir2013effective}. Following \cite{kojima2012school}'s result on the unintended consequences associated with quotas (\cite{abdulkadiroglu2003school}) that limit the number of majority applicants admitted to institutions, \cite{hafalir2013effective} showed that reserves produce significant improvements over quotas. \cite{ehlers2014school} analyzed the constraints imposed by hard upper and lower bounds. To address the limitations of hard reserves, they proposed controlled constraints interpreted as soft bounds. \cite{fragiadakis2017improving} introduced new dynamic quota mechanisms that yield Pareto-superior allocations and respect all distributional constraints. \cite{imamura2025meritocracy} formalized the meritocracy-diversity tradeoff by introducing quantitative measures of meritocracy and diversity for choice rules. Other prominent papers in this category include \cite{dougan2016responsive}, \cite{dougan2022choice}, \cite{celebi2022priority}, \cite{manocha2025gradual}, \cite{celebi2025diversity}, \cite{celebi2025adaptive}, and \cite{dur2024gets}, among others. Our paper departs from this literature by studying the joint implementation of vertical and horizontal reservations in India — a setting heavily regulated by the Supreme Court — within a one-to-all horizontal matching framework under hierarchical horizontal reservations.

The merit-based dominance relation induces a natural partial order over the set of applicants, a concept formalized under various terms in the literature. \cite{arnosti2024explainable} refer to it as "priority-domination," while \cite{sonmez2022affirmative} employ the term "Gale domination." In his seminal work, \cite{gale1968optimal} characterizes a choice that dominates all others with respect to priorities as "optimal." \cite{abdulkadirouglu2021priority} provides a stronger formulation of the priority domination concept introduced by \cite{arnosti2024explainable}. We extend merit-based domination to compare responsive choice functions. \cite{abdulkadirouglu2021priority} also defines a similar comparison for choice rules. Our definition is different from that of \cite{abdulkadirouglu2021priority}.

\section{Conclusion}

This paper provides a framework for analyzing India's affirmative action system, which implements vertical and horizontal reservations simultaneously under a one-to-all horizontal matching policy when horizontal reservations are hierarchical. We introduce the hierarchical choice rule and demonstrate that it is the unique merit-undominated choice rule that satisfies horizontal reservations when these reservations follow a hierarchical structure. We establish that the aggregate choice rule, which comprises hierarchical choice rules across all vertical categories, satisfies substitutability, size monotonicity, and fairness. Our main result shows that the deferred acceptance mechanism coupled with this aggregate choice rule is stable, strategy-proof, and eliminates justified envy.

Although we focus on India's affirmative action system, the theoretical tools developed here have broader applicability. Similar constraint structures arise in various resource allocation problems worldwide, including university admissions with multiple diversity objectives and public housing allocation with layered eligibility criteria. 

\newpage

\section*{APPENDIX A}

\begin{proof}[Proof of Theorem~1]
Fix an applicant set $A$ and write
$D=C^h(A;q^v,\kappa,\rho)$. Throughout, $A_h=A\cap \rho^{-1}(h)$,
and $\succ$ is the strict merit ranking. We use the following
observation: if $i\succ j$, $j\in B$, and $i\notin B$, then
$(B\setminus\{j\})\cup\{i\}\succ_M B$. Moreover, strict merit
domination is transitive on sets of equal cardinality. Indeed,
these assertions follow by matching $i$ with $j$ and every
unchanged applicant with herself, and by composing the
corresponding bijections, respectively.

We process horizontal types sequentially in an order $h_1,\ldots,h_L$ compatible with layers $H_1,\ldots,H_K$, so that each type precedes every type that strictly contains it. Let $E_0=\varnothing$, let $E_\ell$ be the cumulative set of applicants selected after processing $h_1,\ldots,h_\ell$, and put $T_\ell=E_\ell\setminus E_{\ell-1}$. Since distinct types within the same layer have disjoint eligibility sets, sequential processing within a layer produces exactly the selections in the simultaneous construction of $C^h$. In particular, if $m_k=\sum_{j=1}^k|H_j|$ (with $m_0=0$), then $S_k=E_{m_k}$ for $k=0,\ldots,K$. Thus $E_\ell$ refines the cumulative selection process defining $C^h$.

For
$h=h_\ell$,
\[
T_\ell=\operatorname{Top}_{r_h(E_{\ell-1})}
             \bigl(A_h\setminus E_{\ell-1}\bigr),
\qquad
r_h(E_{\ell-1})=
\max\{\kappa_h-|E_{\ell-1}\cap A_h|,0\}.
\]
Finally, let us
\[
T_{L+1}=\operatorname{Top}_{q^v-|E_L|}(A\setminus E_L),
\qquad E_{L+1}=E_L\cup T_{L+1}=D.
\]
Assumption~1 ensures that all these selections respect
capacity.

\textbf{Claim 1.}
For every $\ell\le L$, after processing $h=h_\ell$ we have
\begin{equation}\label{eq:thm1-saturation}
|E_\ell\cap A_h|\ge\min\{\kappa_h,|A_h|\}.
\end{equation}

\begin{proof}[Proof of the Claim 1]
Let $a=|E_{\ell-1}\cap A_h|$. If $a\ge\kappa_h$, the claim is immediate. Otherwise, the
algorithm chooses $\min\{\kappa_h-a,|A_h|-a\}$ additional applicants of type $h$, yielding
$|E_\ell\cap A_h|=\min\{\kappa_h,|A_h|\}$. 
\end{proof}

Selections are never subsequently removed, so the
inequality persists for every previously processed type. The final merit step gives $|D|=\min\{q^v,|A|\}$.
Thus, $D\in\mathcal F(A)$.

If $|A|\le q^v$, then $\mathcal F(A)=\{A\}$ and $D=A$. Henceforth, suppose $|A|>q^v$, so every member of
$\mathcal F(A)$ has cardinality $q^v$. We show that, starting from any $B\in\mathcal F(A)$,
one can reach $D$ by a finite sequence of exchanges, each of which preserves membership in $\mathcal F(A)$ and strictly improves merit.

Inductively, suppose that $B\in\mathcal F(A)$ contains $E_{\ell-1}$, and consider $h=h_\ell$ and $T=T_\ell$. If $T\subseteq B$, no exchange is needed.
Otherwise $T\setminus B\ne\varnothing$. Since $B\in\mathcal F(A)$, this implies $|A_h|>\kappa_h$. If $|A_h|\le\kappa_h$, feasibility would require
$A_h\subseteq B$. Write $E=E_{\ell-1}$ and
$r=\kappa_h-|E\cap A_h|$. Because $T\not\subseteq B$,
we have $r>0$, and because $|A_h|>\kappa_h$,
$|T|=r$. Feasibility of $B$ and $E\subseteq B$ imply
\[
|B\cap(A_h\setminus E)|
=|B\cap A_h|-|E\cap A_h|\ge r=|T|.
\]

Choose $i\in T\setminus B$. Since $T$ consists of the
$r$ highest-merit applicants in $A_h\setminus E$, $\exists j\in\bigl(B\cap(A_h\setminus E)\bigr)\setminus T$,
and $i\succ j$. Set
$\widehat B=(B\setminus\{j\})\cup\{i\}$.
Then $\widehat B\succ_M B$ and $E\subseteq\widehat B$.

\textbf{Claim 2.} 
$\widehat B\in\mathcal F(A)$. 

\begin{proof}[Proof of the Claim 2]
Removing $j$ and adding $i$ does not change the cardinality. For any type $g$ with $j\notin A_{g}$, the number of selected $g$-eligible applicants cannot
decrease. If $j\in A_{g}$, laminarity implies that either
$A_h\subseteq A_{g}$ or $A_{g}\subsetneq A_h$.
In the first case $i\in A_{g}$, so the count is unchanged.
In the second case $g$ precedes $h$ in the chosen order;
by \eqref{eq:thm1-saturation},
$|E\cap A_{g}|\ge\min\{\kappa_g,|A_{g}|\}$, and
$E\subseteq\widehat B$ ensures that the requirement
for $g$ remains satisfied. Thus $\widehat B\in\mathcal F(A)$.
\end{proof}

The exchange also reduces $|T\setminus B|$ by one.
Repeating finitely many times produces a feasible set
containing $E_\ell=E\cup T$, without decreasing merit,
and with a strict improvement whenever an exchange
was necessary. This completes the induction over
$\ell=1,\ldots,L$.

It remains to incorporate the final merit selections.
Take a feasible $B$ containing $E_L$. If $T_{L+1}\subseteq B$,
then $B=D$, since both have cardinality $q^v$.
Otherwise choose $i\in T_{L+1}\setminus B$.
As $|B|=q^v=|E_L|+|T_{L+1}|$, there exists
$j\in B\setminus(E_L\cup T_{L+1})$. By the definition
of $\operatorname{Top}$, $i\succ j$.
Replacing $j$ by $i$ strictly improves merit and
preserves feasibility. $E_L$ remains selected, and
\eqref{eq:thm1-saturation} shows that $E_L$ alone
already meets every horizontal minimum guarantee.
Repeating these exchanges yields $D$.

We have therefore constructed, for every
$B\in\mathcal F(A)$ with $B\ne D$, a finite chain
of feasible sets from $B$ to $D$, containing at least
one strict merit-improving exchange and no
merit-worsening exchange. By transitivity,
\[
D\succ_M B.
\]
In particular, $D$ is merit-undominated within $\mathcal F(A)$. Any other choice rule satisfying horizontal reservations that differs from $C^h$ at some applicant set $A$ selects an element of $\mathcal F(A)$ strictly merit-dominated by $D$. Hence, $C^h$ is the unique merit-undominated choice rule among those satisfying horizontal
reservations.
\end{proof}

\begin{proof}[Proof of Proposition 1]
We first establish the substitutes property by using the sequential representation of the hierarchical choice rule introduced in the proof of Theorem~1.

Fix $A\subseteq I$ and $j\in A$. We compare the execution of the hierarchical choice rule on $A$ with its execution on $A'=A\setminus\{j\}$. Let $E_\ell$ and $E'_\ell$ denote the cumulative sets
selected after stage $\ell$ from $A$ and $A'$,
respectively. In particular,$E_0=E'_0=\emptyset$.

For $\ell=1,\ldots,L$, the selections are determined by
\[
E_\ell
=
E_{\ell-1}\cup
\operatorname{Top}_{r_\ell}
(A_{h_\ell}\setminus E_{\ell-1}),
\]
where
\[
r_\ell
=
\max\{\kappa_{h_\ell}
       -|E_{\ell-1}\cap A_{h_\ell}|,0\}.
\]
The corresponding quantities for $A'$ are denoted
by $r'_\ell$ and $E'_\ell$.

\textit{Claim.}
After every stage $\ell=0,\ldots,L+1$, one of the
following three possibilities holds:
\begin{enumerate}
\item $E'_\ell=E_\ell$,
\item $E'_\ell=E_\ell\setminus\{j\}$, or
\item There exists $z_\ell\in A\setminus E_\ell$
such that $E'_\ell= (E_\ell\setminus\{j\})\cup\{z_\ell\}$.

\end{enumerate}
In the third case, $j\in E_\ell$, and $j$ and $z_\ell$ belong to exactly the same horizontal eligibility sets among the types that remain to
be processed.

\textit{Proof of the claim.}
We proceed by induction on $\ell$. The claim holds at $\ell=0$, since
$E_0=E'_0=\emptyset$. Suppose that it holds after stage $\ell-1$. Write $h=h_\ell$ and consider stage $\ell$.

\medskip
\noindent
\textit{Case 1: $E'_{\ell-1}=E_{\ell-1}$.}

The two executions have identical cumulative
selections. Therefore, their residual quotas for $h$ are
 equal. The sets of applicants available at this stage
differ only in that $j$ is absent from the second
execution.

If $j$ is not among the applicants selected at
this stage in the first execution, the two
executions make identical selections. Hence
$E'_\ell=E_\ell$.

Otherwise, $j$ is selected in the first execution.
Deleting $j$ leaves the other selected applicants
unchanged, because the stage selects applicants
in decreasing order of merit.

If an additional eligible applicant is available,
the second execution selects the highest-merit
such applicant, say $z_\ell$, in place of $j$.
Consequently, $E'_\ell= (E_\ell\setminus\{j\})\cup\{z_\ell\}$.

If no replacement is available, then $E'_\ell=E_\ell\setminus\{j\}$.

In the replacement case, both $j$ and $z_\ell$
belong to $A_h$. Since the horizontal eligibility
sets are laminar and all strict subtypes of $h$
have already been processed, every remaining
horizontal type either contains $A_h$ or is
disjoint from $A_h$.

Thus $j$ and $z_\ell$ have identical eligibility
for every remaining horizontal type.

\medskip
\noindent
\textit{Case 2:
$E'_{\ell-1}=E_{\ell-1}\setminus\{j\}$.}

In this case $j$ has already been selected in the first execution, so it is not available for selection at stage $\ell$ in either execution. The remaining applicants for type $h$ are identical.

If $j\notin A_h$, the residual quotas are equal,
and both executions select the same applicants.
The difference between their cumulative selections
remains exactly $\{j\}$.

If $j\in A_h$, the second execution has one fewer
previously selected $h$-eligible applicant. Thus, its residual quota is one greater whenever the first execution's residual quota is positive. When that residual quota is zero, the second execution may have a residual quota
of one. Consequently, the second execution selects all applicants selected at this stage in the first execution and, if enough eligible
applicants remain, one additional applicant
$z_\ell$.

If such an applicant is selected, then $E'_\ell=(E_\ell\setminus\{j\})\cup\{z_\ell\}$. Otherwise, $E'_\ell=E_\ell\setminus\{j\}$.

In the former case, $j,z_\ell\in A_h$, so
laminarity and the processing order again imply
identical eligibility for all remaining types.

\medskip
\noindent
\textit{Case 3:
$E'_{\ell-1}
=(E_{\ell-1}\setminus\{j\})\cup\{z\}$,}
where $z\notin E_{\ell-1}$.

By the induction hypothesis, $j$ and $z$
have identical eligibility for every remaining
horizontal type. In particular,
\[
j\in A_h
\quad\Longleftrightarrow\quad
z\in A_h.
\]

Since the second cumulative selection replaces
$j$ with $z$, the two executions have the same
number of previously selected $h$-eligible
applicants. Thus, $r'_\ell=r_\ell$.

If $j,z\notin A_h$, the available applicant pools
and residual quotas are identical. Both executions
make the same new selections, and the replacement
relationship persists.

Now suppose $j,z\in A_h$. Applicant $j$ has
already been selected in the first execution,
while $z$ has already been selected in the second.
Thus, the applicant pool in the second execution
is obtained from the first execution's pool by
deleting $z$.

Both executions select the same number of
applicants, up to the number available.

If $z$ is not selected at this stage in the first
execution, the two executions select exactly the
same new applicants. The replacement relationship
therefore persists.

If $z$ is selected at this stage in the first
execution, the second execution selects every
other applicant selected in the first execution.
It also selects the next-highest-merit available
applicant, say $w$, if one exists.

If such a replacement exists, then $E'_\ell
=(E_\ell\setminus\{j\})\cup\{w\}$. Otherwise, $E'_\ell=E_\ell\setminus\{j\}$.

In the former case, $j,w\in A_h$, and the same
laminarity argument establishes identical
eligibility for every remaining horizontal type.

This completes the induction for all horizontal
stages $\ell=1,\ldots,L$.

Finally, consider the merit-based stage $L+1$.
At this stage every remaining applicant is
eligible, and the number of available positions
is determined by the remaining capacity
$q_v-|E_L|$.

The preceding three cases apply with the
eligibility set equal to $I$ and the quota
equal to $q_v$. Since the stage selects the
highest-merit remaining applicants, deleting
$j$ either leaves the final selection unchanged,
removes $j$ without replacement, or replaces
$j$ by one additional applicant.

Therefore, $E'_{L+1} \supseteq E_{L+1}\setminus\{j\}$. This proves the claim.

We now establish substitutability. By the definition of the hierarchical choice rule, $E_{L+1}=C^h(A;q_v,\kappa,\rho)$ and $E'_{L+1}=C^h(A\setminus\{j\};q_v,\kappa,\rho)$.

The claim therefore implies $C^h(A;q_v,\kappa,\rho)\setminus\{j\}
\subseteq C^h(A\setminus\{j\};q_v,\kappa,\rho)$.

Now take any $B\subseteq I$ and distinct
$i,j\notin B$. If
\[
i\in C^h(B\cup\{i,j\};q_v,\kappa,\rho),
\]
then applying the above claim to $A=B\cup\{i,j\}$
gives
\[
i\in C^h(B\cup\{i\};q_v,\kappa,\rho).
\]

Hence $C^h$ satisfies the substitutes property.

\medskip
We next establish size monotonicity.

By Assumption~1, the horizontal selection stages
never select more than $q_v$ applicants.
The final stage fills all remaining positions
by merit whenever sufficiently many applicants
are available. Consequently, for every
$A\subseteq I$,
\[
|C^h(A;q_v,\kappa,\rho)|
=
\min\{q_v,|A|\}.
\]

For any $A\subseteq B\subseteq I$,
\[
\begin{aligned}
|C^h(A;q_v,\kappa,\rho)|=\min\{q_v,|A|\} \leq\min\{q_v,|B|\}=|C^h(B;q_v,\kappa,\rho)|.
\end{aligned}
\]

Therefore, $C^h$ satisfies size monotonicity.
\end{proof}

\begin{proof}[Proof of Theorem 2]
We need to show that $\mathcal{C}^{h}$ is fair. Toward a contradiction, suppose not. Then, there exist a set of contracts $Y\subset X_{s}$ and a pair of contracts $x,y\in Y$ such that $x\notin \mathcal{C}^{h}(Y)$, $y\in \mathcal{C}^{h}(Y)$, $\mathbf{i}(x)\succ_{s}\mathbf{i}(y)$ and $\mathbf{t}(x)=\mathbf{t}(y)=v$, and $\rho(\mathbf{i}(x))\supseteq\rho(\mathbf{i}(y))$. 

Given $x\notin C^{h}_{v}(Y)$ and $y\in C^{h}_{v}(Y)$, let $
Z=(C_{v}^{h}(Y^{'})\setminus\{y\})\cup\{x\},$ where $Y^{'}\subseteq Y$ is the set of contracts considered by category $v$. $\mathbf{i}(Z)$ merit-dominates $\mathbf{i}(C_{v}^{h}(Y^{'}))$ because $\mathbf{i}(x)$ has a higher score than $\mathbf{i}(y)$, contradicting $C^{h}$ being merit-undominated. Thus, $\mathcal{C}^{h}$ is fair. 
\end{proof}

\begin{proof}[Proof of Proposition 2]
Let $S=\{s\}$ and $I=\{i,j\}$. Institution $s$ has one open category position and one category-$r$ position. Both applicants are eligible for category $r$, neither has a horizontal type, and $i\succ_s j$. Write $x_o=(i,s,o)$, $x_r=(i,s,r)$, $y_o=(j,s,o)$, and $y_r=(j,s,r)$. For every set of offered contracts, let $C_s$ first select the highest-merit applicant offering an open category contract, if any, and then select the highest-merit remaining applicant offering a category-$r$ contract, if any. No applicant is selected twice. This choice rule is fair.

First, let $i$ find $(s,r)$ acceptable and let $j$ find $(s,o)$ acceptable. Consider $Y=\{x_r\}$. Applicant $j$ cannot have justified envy toward $i$, because $i\succ_s j$, and no other applicant is assigned. Thus $Y$ eliminates justified envy. However, $C_s(Y\cup\{y_o\})=\{x_r,y_o\}$, so $\{y_o\}$ blocks $Y$. Hence $Y$ is not stable (Definition 7, condition (3)).

Second, suppose $(s,r)\;P_i\;(s,o)\;P_i\;\varnothing$ and $(s,r)\;P_j\;\varnothing\;P_j\;(s,o)$. Consider $Z=\{x_o,y_r\}$. Both contracts are individually rational, and $C_s(Z)=Z$. Applicant $j$ has no preferred alternative contract, and the only contract preferred by $i$ to $x_o$ is $x_r$. But $C_s(Z\cup\{x_r\})=Z$: the open position is assigned to $i$ first, and the category-$r$ position to $j$. Consequently, no set of contracts blocks $Z$, and $Z$ is stable. Nevertheless, $i$ prefers $j$'s category-$r$ position to her own open position, has higher merit, and has the same horizontal eligibility as $j$. Thus $Z$ does not eliminate justified envy. The two properties are independent.
\end{proof}

\begin{proof}[Proof of Theorem 4]

We first show that $\mathcal{C}^{hT}$ is fair. Toward a contradiction, suppose not. Then, there exists a set of contracts $Y\subseteq X_{s}$ and contracts $x,y\in Y$ such that $x\notin \mathcal{C}^{hT}(Y)$, $y\in \mathcal{C}^{hT}(Y)$, $\mathbf{i}(x)\succ_{s}\mathbf{i}(y)$, $\mathbf{t}(x)=\mathbf{t}(y)=v$, and $\rho(\mathbf{i}(x))\supseteq\rho(\mathbf{i}(y))$.

$x\notin \mathcal{C}^{hT}$ implies $x\notin C_{v^{}}^{h}$ and $x\notin C_{D}$. Note that $y$ cannot be chosen in category-D while $x$ is rejected because the choice rule $C_{D}$ is responsive, induced by the score rankings, and $\mathbf{i}(x)\succ_{s}\mathbf{i}(y)$. Then, must be the case that $x,y\in Y^{'}\subseteq Y$, $x\notin C_{v}^{h}(Y^{'})$ and $y\in C_{v}^{h}(Y^{'})$, where $Y^{'}$ is the set of contracts category-$v$ receives. 

Let $Z=(C_{v}^{h}(Y^{'})\setminus\{y\})\cup\{x\}.$ Note that $\mathbf{i}(Z)$ merit-dominates $\mathbf{i}(C_{v}^{h}(Y^{'}))$ because $\mathbf{i}(x)$ has a higher score than $\mathbf{i}(y)$, contradicting $C_{v}^{h}$ is merit-undominated. Thus, $\mathcal{C}^{hT}$ is fair.
\end{proof}

\begin{proof}[Proof of Proposition 4]
(i) Let $\sigma$ be such that $B\in\mathcal{F}^{1}(A)$. The sets $\sigma^{-1}(h)$, $h\in H$, are pairwise disjoint subsets of $B\cap A_H$, and each has at least $\min\{\kappa_h,|A_h|\}$ elements.

(ii) Let $r=\max\{\kappa_h,\kappa_{h'}\}$. Every $B\in F(A)$ satisfies $|B\cap A_h|\geq\kappa_h$ and, because $A_{h'}\subseteq A_h$, also $|B\cap A_h|\geq\kappa_{h'}$. Hence $|B\cap A_h|\geq r$. To attain the bound, select $\kappa_{h'}$ applicants from $A_{h'}$ and then select $r-\kappa_{h'}$ further applicants from $A_h$ not already selected. There are enough applicants in $A_h$, because $|A_h|\geq\kappa_h+\kappa_{h'}\geq r$. Fill the remaining $q-r$ positions with applicants in $A\setminus A_h$. The resulting set is in $F(A)$ and contains exactly $r$ members of $A_h$.

Under one-to-one counting, $\sigma^{-1}(h)$ and $\sigma^{-1}(h')$ are disjoint subsets of $B\cap A_h$ with cardinalities at least $\kappa_h$ and $\kappa_{h'}$, respectively. Thus $|B\cap A_h|\geq\kappa_h+\kappa_{h'}$. To attain this bound, select $\kappa_{h'}$ applicants from $A_{h'}$ and assign them to $h'$, then select $\kappa_h$ distinct additional applicants from $A_h$ and assign them to $h$. This is possible because $|A_h|\geq\kappa_h+\kappa_{h'}$. Fill the remaining $q-\kappa_h-\kappa_{h'}$ positions with applicants outside $A_h$; enough such applicants exist because $q-\kappa_h-\kappa_{h'}\leq q-r\leq|A\setminus A_h|$. The assumption $q\geq\kappa_h+\kappa_{h'}$ ensures that the number of remaining positions is nonnegative. The two minima differ strictly because both quotas are positive. The one-to-all construction cannot satisfy the one-to-one requirements, proving strict inclusion.
\end{proof}

\begin{proof}[Proof of Proposition 5]
Let $q=2$, $\kappa_{h_1}=\kappa_{h_2}=1$, and let all other horizontal reservations be zero. Pick individuals $a\in\rho^{-1}(h_1)\cap\rho^{-1}(h_2)$, $b\in\rho^{-1}(h_1)\setminus\rho^{-1}(h_2)$, $c\in\rho^{-1}(h_2)\setminus\rho^{-1}(h_1)$, and $d$ with $\rho(d)\cap\{h_1,h_2\}=\emptyset$, and suppose the merit ranking is $d\succ a\succ b\succ c$.

(i) From $A=\{a,b,c,d\}$ the reservation-feasible sets of size two are $\{a,d\}$, $\{a,b\}$, $\{a,c\}$ and $\{b,c\}$. Note that $\{a,d\}$ merit-dominates each of the other three. Hence any choice rule $C$ that satisfies horizontal reservations and is merit-undominated has $C(A)=\{a,d\}$. From $A'=\{b,c,d\}$ the unique reservation-feasible set is $\{b,c\}$, so $C(A')=\{b,c\}$. Thus $d\in C(A'\cup\{a\})$ while $d\notin C(A')$, which violates the substitutes property. The two individuals $a$ and $d$ are complements. Note that $a$ alone satisfies both reservations and thereby frees a position for $d$.

(ii) Let $S=\{s,s'\}$ and $I=\{a,b,c,d\}$, with a single vertical category so that a contract is a pair $(i,s)$. Institution $s$ has the capacity and reservations of part (i) and a choice rule $C_s$ that satisfies horizontal reservations and is merit-undominated. For every $A\subseteq\{a,b,c,d\}$ the family $\mathcal{F}(A)$ has a unique merit-undominated element, so $C_s$ is uniquely determined, and in particular
\[
C_s(\{a,b,c,d\})=\{a,d\},\quad C_s(\{a,b,d\})=C_s(\{a,c,d\})=\{a,d\} 
\]
\[
C_s(\{a,b,c\})=\{a,b\},\quad C_s(\{b,c,d\})=\{b,c\},
\]

while $C_s(A)=A$ whenever $|A|\le2$. Institution $s'$ has one position, no horizontal reservations, and merit ranking $d\succ_{s'}a$. The choice rule of institution $s'$ is responsive. The preferences are $P_a: s'\,,\,s\,,\,\emptyset$ and $P_d: s\,,\,s'\,,\,\emptyset$. Individuals $b$ and $c$ find only $s$ acceptable.

Suppose $Y$ is a stable matching. We distinguish cases by $a$'s assignment.

\emph{Case 1: $a\in i(Y_{s'})$.} Then $Y_{s'}=\{(a,s')\}$ and $i(Y_s)\subseteq\{b,c,d\}$. If $d\notin i(Y_s)$, then $d$ is unmatched, and $\{(d,s')\}$ blocks $Y$ because $C_{s'}(\{(a,s'),(d,s')\})=\{(d,s')\}$. If $d\in i(Y_s)$, condition (2) of stability rules out $i(Y_s)=\{b,c,d\}$ since $C_s(\{b,c,d\})=\{b,c\}$, so $i(Y_s)$ is one of $\{d\}$, $\{b,d\}$, $\{c,d\}$. In each case an individual in $\{b,c\}\setminus i(Y_s)$ is unmatched and blocks with $s$: $b \in C_s(\{b,d\})$, $c \in C_s(\{b,c,d\})$ and $b \in C_s(\{b,c,d\})$, respectively.

\emph{Case 2: $a\in i(Y_s)$.} If $d\in i(Y_s)$ then $Y_s=\{(a,s),(d,s)\}$ and $s'$ holds no contract. Since $a$ prefers $s'$ to $s$ and $C_{s'}(\{(a,s')\})=\{(a,s')\}$, the contract $(a,s')$ blocks $Y$. If $d\notin i(Y_s)$, then $i(Y_s)$ is one of $\{a\}$, $\{a,b\}$, $\{a,c\}$, and $d$, who is either at $s'$ or unmatched, prefers $s$. Because $d\in C_s(i(Y_s)\cup\{d\})$ in all three cases, $(d,s)$ blocks $Y$.

\emph{Case 3: $a$ is unmatched.} If $d\notin i(Y_{s'})$, then $(a,s') \in C_{s'}(Y_{s'}\cup\{(a,s')\})$ and $(a,s')$ blocks $Y$. If $d\in i(Y_{s'})$, then $i(Y_s)\subseteq\{b,c\}$, and $a\in C_s(i(Y_s)\cup\{a\})$ in every case ($C_s(\{a,b,c\})=\{a,b\}$, and $C_s(A)=A$ for $|A|\le2$), so $(a,s)$ blocks $Y$.

No case admits a stable matching.
\end{proof}

\section*{APPENDIX B}

To prove Theorems \ref{mechanism} and \ref{extensionresult2}, we first prove a technical result, Proposition \ref{technical}, based mainly on the theory introduced in \cite{hatfield2017restud} and \cite{hatfield2017stable}. 

\begin{definition}[\cite{hatfield2017restud}]
    An \textbf{offer process} for institution $s$, with the aggregate choice rule $C_{s},$ is a finite sequence of distinct contracts $(x^{1},x^{2},...,x^{M})$, such that for all $m=1,...,M$, $x^{m}\in X_{s}$. An offer process $(x^{1},x^{2},...,x^{M})$ for $s$ is $\textbf{observable}$ if, for all $m=1,...,M$, $\mathbf{i}(x^{m})\notin\mathbf{i}(C_{s}(\{x^{1},...,x^{m-1}\}))$. 
\end{definition}

We adopt the following notation used in \cite{hatfield2017restud}.

If $X^{M}=\{x^{1},...,x^{M}\}$ is an observable offer process, we say $X^{m}=\{x^{1},..,x^{m}\}$. That is, $X^{m}$ are the contracts proposed up to step $m$ of the observable offer process $X^{M}$. 

Let $H_{k}(X^{m})$ be the set of contracts available for category $k$ in the calculation of $C_{s}(X^{m})$. 

$F_{k}(X^{m})=\underset{n\leq m}{\cup}H_{k}(X^{n})$. That is, $F_{k}(X^{m})$ is the set of all contracts available to category $k$ at some point in the offer process $X^{m}=\{x^{1},...,x^{m}\}$. 

$q_{k}^{m-1}(r_{1},...,r_{k-1})$ and $q_{k}^{m}(\widetilde{r}_{1},...,\widetilde{r}_{k-1})$ are the dynamic capacity of category $k$ in Steps $m-1$ and $m$ of the observable offer process,respectively. 

$R_{k}(X^{m};q_{k}^{m}(\widetilde{r}_{1},...,\widetilde{r}_{k-1}))$ is the set of contracts rejected by category $k$ at step $m$ of the observable offer process $X^{m}$. 

The following Proposition is the adaptation of Claim 1 in \cite{hatfield2017stable} to the generalized sequential (GS) choice rules introduced in \cite{aygun2026matching}. The proof idea is the same as that in \cite{hatfield2017stable}, even though GS choice rules are defined differently from the choice rules with flexible allotments in \cite{hatfield2017stable}.

\begin{proposition}\label{technical}
For all categories $k\in\{1,...,K_{s}\}$ and for all $m\in\{1,...,M\}$
where $M$ is the last step of observable offer process $X^{M}=\{x^{1},...,x^{M}\}$: 

1. $C_{k}(H_{k}(X^{m-1});q_{k}^{m-1}(r_{1},...,r_{k-1}))\subseteq H_{k}(X^{m})$.

2. $C_{k}(F_{k}(X^{m});q_{k}^{m}(\widetilde{r}_{1},...,\widetilde{r}_{k-1}))\subseteq C_{k}(H_{k}(X^{m-1});q_{k}^{m-1}(r_{1},...,r_{k-1}))\cup[H(X^{m})\setminus H_{k}(X^{m-1})]$.

3. $C_{k}(H_{k}(X^{m});q_{k}^{m}(\widetilde{r}_{1},...,\widetilde{r}_{k-1}))=C_{k}(F_{k}(X^{m});q_{k}^{m}(\widetilde{r}_{1},...,\widetilde{r}_{k-1}))$.

4. $q_{k}^{m-1}(r_{1},...,r_{k-1})\geq q_{k}^{m}(\widetilde{r}_{1},...,\widetilde{r}_{k-1})$.\footnote{This result is specific to GS choice rules. There is no counterpart of this condition in \cite{hatfield2017stable}.} 

5. $R_{k}(F_{k}(X^{m-1});q_{k}^{m-1}(r_{1},...,r_{k-1}))\subseteq R_{k}(F_{k}(X^{m});q_{k}^{m}(\widetilde{r}_{1},...,\widetilde{r}_{k-1}))$.
\end{proposition}

\paragraph{Proof}

We use mathematical induction on pairs $(m,k)$ as done in \cite{hatfield2017stable}.
\[
(1,1),(1,2),...,(1,K),(2,1),(2,2),...,(2,K),...,(M,1),(M,2),...,(M,K).
\]

\paragraph{Initial Step:}

Consider $m=1$ and any category $k=1,...,K$. Note that $X^{m-1}=X^{0}=\emptyset$ and $X^{1}=\{x^{1}\}$. Since $H_{k}(X^{0})=H_{k}(\emptyset)=\emptyset$, condition (1) holds trivially because $\emptyset\subseteq H_{k}(X^{1})$ for all $k=1,...,K$. Condition (2) also holds because it reduces to $C_{k}(F_{k}(X^{1})\subseteq H_{k}(X^{1})=F_{k}(X^{1})$. Condition (3) also holds trivially since $H_{k}(X^{1})=F_{k}(X^{1})$. Condition (4) holds for the pair $(1,1)$ as for the first category the initial capacity is given exogenously. Condition (5) reduces to $R_{k}(\emptyset)=\emptyset\subseteq R_{k}(F_{k}(X^{1}))$ and is trivially true. 

\paragraph{Inductive assumption:}

Assume that conditions (1)-(5) hold for 
\begin{itemize}
\item every $(m^{'},k)$ with $m^{'}<m$ and $k=1,...,K$,
\item every $(m,k^{'})$ with $k^{'}<k$. 
\end{itemize}
We need to show that conditions (1)-(5) hold for the pair $(m,k)$, starting with (1). 

\paragraph{(1)}
Take $z\in C_{k}(H_{k}(X^{m-1});q_{k}^{m-1}(r_{1},...,r_{k-1}))$. If $z$ is chosen by category $k$, then it must have been rejected by all the categories that precede it. Hence, we have 
\[
(\{x^{1},...,x^{m-1}\})_{\mathbf{i}(z)}\subseteq \underset{k^{'}<k}{\bigcap}R_{k^{'}}(H_{k^{'}}(X^{m-1})).
\]

By inductive assumptions (2) and (3), for all $k^{'}<k$, we have
\[
C_{k^{'}}(H_{k^{'}}(X^{m});q_{k^{'}}^{m}(\widetilde{r}_{1},...,\widetilde{r}_{k^{'}-1}))\subseteq C_{k^{'}}(H_{k^{'}}(X^{m-1});q_{k^{'}}^{m-1}(r_{1},...,r_{k}))\cup[H_{k^{'}}(X^{m})\setminus H_{k^{'}}(X^{m-1})].
\]

 Note that all the contracts of agent $\mathbf{i}(z)$ are in $H_{k^{'}}(X^{m-1})$ for all $k^{'}<k$. Hence, $\mathbf{i}(z)\notin\mathbf{i}[H_{k^{'}}(X^{m})\setminus H_{k^{'}}(X^{m-1})]$. We also know that $\mathbf{i}(z)\notin\mathbf{i}[C_{k^{'}}(H_{k^{'}}(X^{m-1});q_{k^{'}}^{m-1}(r_{1},...,r_{k}))]$ because $z$ is chosen by category $k$ in the offer process $X^{m-1}$.
Then we have 
\[
z\notin C_{k^{'}}(H_{k^{'}}(X^{m});q_{k^{'}}^{m}(\widetilde{r}_{1},...,\widetilde{r}_{k^{'}-1})),\;\forall k^{'}<k.
\]

This means that $z$ is not chosen by any category that precedes category
$k$. In other words, $(\{x^{1},...,x^{m-1}\})_{\mathbf{i}(z)}\subseteq\ \underset{k^{'}<k}{\bigcap} R_{k^{'}}(H_{k^{'}}(X^{m}))$.
Therefore, we have $z\in H_{k}(X^{m})$. 

\paragraph{(4)}
By inductive assumption, (4) holds for $(i)$ every $(m^{'},k)$ with $m^{'}<m$ and $k=1,...,K$, and $(ii)$ every $(m,k^{'})$ with $k^{'}<k$. To show that $q_{k}^{m-1}(r_{1},...,r_{k-1})\geq q_{k}^{m}(\widetilde{r}_{1},...,\widetilde{r}_{k-1})$, we first compare $r_{k^{'}}^{m-1}$ and $\widetilde{r}_{k^{'}}^{m}$.
By definition, 
\[
r_{k^{'}}^{m-1}=q_{k^{'}}^{m-1}(r_{1},...,r_{k^{'}-1})-\mid C_{k^{'}}(H_{k^{'}}(X^{m-1});q_{k^{'}}^{m-1}(r_{1},...,r_{k^{'}-1}))\mid.
\]

By inductive assumption, we have $q_{k^{'}}^{m-1}(r_{1},...,r_{k^{'}-1})\geq q_{k^{'}}^{m}(\widetilde{r}_{1},...,\widetilde{r}_{k^{'}-1})$.
We combine it with the inequality above and get 
\[
r_{k^{'}}^{m-1}\geq q_{k^{'}}^{m}(\widetilde{r}_{1},...,\widetilde{r}_{k^{'}-1})-\mid C_{k^{'}}(H_{k^{'}}(X^{m-1});q_{k^{'}}^{m-1}(r_{1},...,r_{k^{'}-1}))\mid.
\]

By inductive assumption (3), for all $k^{'}<k$, we have 
\[
C_{k^{'}}(H_{k^{'}}(X^{m-1});q_{k^{'}}^{m-1}(\widetilde{r}_{1},...,\widetilde{r}_{k^{'}-1}))=C_{k^{'}}(F_{k^{'}}(X^{m-1});q_{k^{'}}^{m-1}(\widetilde{r}_{1},...,\widetilde{r}_{k^{'}-1})).
\]

By size monotonicity of the choice functions and the fact
that $F_{k^{'}}(X^{m-1})\subseteq F_{k^{'}}(X^{m})$, 
\[
\mid C_{k^{'}}(F_{k^{'}}(X^{m-1});q_{k^{'}}^{m-1}(r_{1},...,r_{k^{'}-1}))\mid\leq\mid C_{k^{'}}(F_{k^{'}}(X^{m});q_{k^{'}}^{m-1}(r_{1},...,r_{k^{'}-1}))\mid.
\]

Hence, 
\[
r_{k^{'}}^{m-1}\geq q_{k^{'}}^{m}(\widetilde{r}_{1},...,\widetilde{r}_{k^{'}-1})-\mid C_{k^{'}}(F_{k^{'}}(X^{m-1});q_{k^{'}}^{m-1}(r_{1},...,r_{k^{'}-1}))\mid.
\]

By the quota monotonicity of categories' choice rule, we have 
$$
\mid C_{k^{'}}(F_{k^{'}}(X^{m});q_{k^{'}}^{m-1}(r_{1},...,r_{k^{'}-1}))\mid-\mid C_{k^{'}}(F_{k^{'}}(X^{m});q_{k^{'}}^{m}(\widetilde{r}_{1},...,\widetilde{r}_{k^{'}-1}))\mid \leq 
$$

$$q_{k^{'}}^{m-1}(r_{1},...,r_{k^{'}-1})-q_{k^{'}}^{m}(\widetilde{r}_{1},...,\widetilde{r}_{k^{'}-1}).
$$

Rearranging the terms gives 
\[
q_{k^{'}}^{m-1}(r_{1},...,r_{k^{'}-1})-\mid C_{k^{'}}(F_{k^{'}}(X^{m});q_{k^{'}}^{m-1}(r_{1},...,r_{k^{'}-1}))\mid\geq 
\]
\[
q_{k^{'}}^{m}(\widetilde{r}_{1},...,\widetilde{r}_{k^{'}-1})-\mid C_{k^{'}}(F_{k^{'}}(X^{m});q_{k^{'}}^{m}(\widetilde{r}_{1},...,\widetilde{r}_{k^{'}-1}))\mid=\widetilde{r}_{k^{'}}^{m}.
\]

Combining the last inequality with $r_{k^{'}}^{m-1}\geq q_{k^{'}}^{m}(\widetilde{r}_{1},...,\widetilde{r}_{k^{'}-1})-\mid C_{k^{'}}(F_{k^{'}}(X^{m-1});q_{k^{'}}^{m-1}(r_{1},...,r_{k^{'}-1}))\mid$
gives $r_{k^{'}}^{m-1}\geq\widetilde{r}_{k^{'}}^{m}$ for
all $k^{'}=1,...,k-1$. Hence, by the monotonicity of the capacity
transfers we conclude that $$q_{k}^{m-1}(r_{1},...,r_{k-1})\geq q_{k}^{m}(\widetilde{r}_{1},...,\widetilde{r}_{k-1}).$$ 

\paragraph{(5)}
By (4), we know that $q_{k}^{m-1}(r_{1},...,r_{k-1})\geq q_{k}^{m}(\widetilde{r}_{1},...,\widetilde{r}_{k-1})$.
By the quota monotonicity,
\[
C_{k}(F_{k}(X^{m-1});q_{k}^{m}(\widetilde{r}_{1},...,\widetilde{r}_{k-1}))\subseteq C_{k}(F_{k}(X^{m-1});q_{k}^{m-1}(r_{1},...,r_{k-1})).
\]

Hence, we have
\[
F_{k}(X^{m-1})\setminus C_{k}(F_{k}(X^{m-1});q_{k}^{m}(\widetilde{r}_{1},...,\widetilde{r}_{k-1}))\supseteq
\]
\[
F_{k}(X^{m-1})\setminus C_{k}(F_{k}(X^{m-1});q_{k}^{m-1}(r_{1},...,r_{k-1}))=R_{k}(F_{k}(X^{m-1});q_{k}^{m-1}(r_{1},...,r_{k-1})).
\]

Then, by the substitutes property and the fact that
$F_{k}(X^{m-1})\subseteq F_{k}(X^{m})$, 
\[
R_{k}(F_{k}(X^{m});q_{k}^{m}(\widetilde{r}_{1},...,\widetilde{r}_{k-1}))=F_{k}(X^{m})\setminus C_{k}(F_{k}(X^{m});q_{k}^{m}(\widetilde{r}_{1},...,\widetilde{r}_{k-1}))\supseteq
\]

\[
F_{k}(X^{m-1})\setminus C_{k}(F_{k}(X^{m-1});q_{k}^{m}(\widetilde{r}_{1},...,\widetilde{r}_{k-1})).
\]

Hence, we conclude that 
\[
R_{k}(F_{k}(X^{m-1});q_{k}^{m-1}(r_{1},...,r_{k-1}))\subseteq R_{k}(F_{k}(X^{m});q_{k}^{m}(\widetilde{r}_{1},...,\widetilde{r}_{k-1})).
\]

\paragraph{(2)}
By the definition of choice rules, we have $C_{k}(F_{k}(X^{m});q_{k}^{m}(\widetilde{r}_{1},...,\widetilde{r}_{k-1}))\subseteq F_{k}(X^{m})$.
We can decompose $F_{k}(X^{m})$ as follows: 
\[
F_{k}(X^{m})=(F_{k}(X^{m}))\setminus F_{k}(X^{m-1}))\cup[R_{k}(F_{k}(X^{m-1}))\cup C_{k}(F_{k}(X^{m-1});q_{k}^{m-1}(r_{1},...,r_{k-1}))],
\]

since $F_{k}(X^{m-1})=R_{k}(F_{k}(X^{m-1}))\cup C_{k}(F_{k}(X^{m-1});q_{k}^{m-1}(r_{1},...,r_{k-1}))$
by definition. If we replace $R_{k}(F_{k}(X^{m-1}))$ by $R_{k}(F_{k}(X^{m}))$
and use the fact that $R_{k}(F_{k}(X^{m-1}))\subseteq R_{k}(F_{k}(X^{m}))$
by (5), then we have 
\[
C_{k}(F_{k}(X^{m});q_{k}^{m}(\widetilde{r}_{1},...,\widetilde{r}_{k-1}))\subseteq(F_{k}(X^{m}))\setminus F_{k}(X^{m-1}))\cup[R_{k}(F_{k}(X^{m}))\cup C_{k}(F_{k}(X^{m-1});q_{k}^{m-1}(r_{1},...,r_{k-1}))].
\]

But, by definition, $C_{k}(F_{k}(X^{m});q_{k}^{m}(\widetilde{r}_{1},...,\widetilde{r}_{k-1})\cap R_{k}(F_{k}(X^{m}))=\emptyset$.
Hence, the above inclusion relation can be written as 
\[
C_{k}(F_{k}(X^{m});q_{k}^{m}(\widetilde{r}_{1},...,\widetilde{r}_{k-1}))\subseteq(F_{k}(X^{m}))\setminus F_{k}(X^{m-1}))\cup C_{k}(F_{k}(X^{m-1});q_{k}^{m-1}(r_{1},...,r_{k-1})).
\]

By the definition of $F_{k}(X^{m})$ and $F_{k}(X^{m-1})$, we know
that 
\[
F_{k}(X^{m})\setminus F_{k}(X^{m-1})\equiv H_{k}(X^{m})\setminus\cup_{n<m}H_{k}(X^{n})\subseteq H_{k}(X^{m})\setminus H_{k}(X^{m-1}).
\]

Then, we can conclude that 
\[
C_{k}(F_{k}(X^{m});q_{k}^{m}(\widetilde{r}_{1},...,\widetilde{r}_{k-1}))\subseteq(H_{k}(X^{m}))\setminus H_{k}(X^{m-1}))\cup C_{k}(F_{k}(X^{m-1});q_{k}^{m-1}(r_{1},...,r_{k-1})).
\]

Finally, by the inductive assumption $(3)$, we have 
\[
C_{k}(H_{k}(X^{m-1});q_{k}^{m-1}(r_{1},...,r_{k-1}))=C_{k}(F_{k}(X^{m-1});q_{k}^{m-1}(r_{1},...,r_{k-1})).
\]

Hence, we can conclude that 
\[
C_{k}(F_{k}(X^{m});q_{k}^{m}(\widetilde{r}_{1},...,\widetilde{r}_{k-1}))\subseteq(H_{k}(X^{m}))\setminus H_{k}(X^{m-1}))\cup C_{k}(H_{k}(X^{m-1});q_{k}^{m-1}(r_{1},...,r_{k-1})).
\]

\paragraph{(3)}
By $(2)$ we have that 
\[
C_{k}(F_{k}(X^{m});q_{k}^{m}(\widetilde{r}_{1},...,\widetilde{r}_{k-1}))\subseteq(H_{k}(X^{m}))\setminus H_{k}(X^{m-1}))\cup C_{k}(H_{k}(X^{m-1});q_{k}^{m-1}(r_{1},...,r_{k-1})).
\]

By $(1)$ we have $C_{k}(H_{k}(X^{m-1});q_{k}^{m-1}(r_{1},...,r_{k-1}))\subseteq H_{k}(X^{m})$.
Then, combining $(1)$ and $(2)$ gives us 
\[
C_{k}(F_{k}(X^{m});q_{k}^{m}(\widetilde{r}_{1},...,\widetilde{r}_{k-1}))\subseteq H_{k}(X^{m}).
\]

Hence, since the choice rules satisfy the IRC, we conclude that 
\[
C_{k}(H_{k}(X^{m});q_{k}^{m}(\widetilde{r}_{1},...,\widetilde{r}_{k-1}))=C_{k}(F_{k}(X^{m});q_{k}^{m}(\widetilde{r}_{1},...,\widetilde{r}_{k-1})).
\]

\begin{proof}[Proof of Theorem 3]
We borrow the following definition from  \cite{aygun2026matching}.

\begin{definition}[\cite{aygun2026matching}]
An aggregate choice rule $\mathcal{C}$ is a \textbf{GSq rule} if it can be represented as a pair $\left(\left(C_{v}\right)_{v\in V},q\right)$ of a collection of category choice rules and a transfer policy, where each category's choice rule satisfies the substitutes property, size monotonicity, and quota monotonicity, and the transfer policy satisfies monotonicity.
\end{definition}

We prove Theorem 3 by showing that the aggregate choice rule $\mathcal{C}^{h}$ is in the GS family by using the following result.

\begin{corollary}[\cite{aygun2026matching}]
    The COM is the unique mechanism that is stable and strategy-proof when institutions' choice rules are in the GSq family. 
\end{corollary}

 Thus, we need to show that the hierarchical choice rule $C^{h}$ satisfies the substitutes property, size monotonicity, and quota monotonicity property. Note that since there is no capacity transfer in this case, monotonicity of transfer policy is trivially satisfied. We have already shown that $C^{h}$ satisfies the substitutes property and size monotonicity. We just need to show that the hierarchical choice rule $C^{h}$ satisfies the quota monotonicity. 

\begin{definition}[\cite{aygun2026matching}]
A choice rule $C_{s}^{k}: 2^{X} \times\mathbb{Z}_{+}\longrightarrow2^{X_{s}}$ satisfies \text{quota monotonicity} (QM) if for any $q\in\mathbb{\mathbb{Z}}_{+}$ such that for all $Y\subseteq X$,

\begin{itemize}
    \item $C_{s}^{v}(Y,q)\subseteq C_{s}^{v}(Y,1+q)$, and
    \item $\mid C_{s}^{v}(Y,1+q)\mid-\mid C_{s}^{v}(Y,q)\mid\leq 1$.
\end{itemize}
\end{definition}

Consider a set of contracts $Y\subseteq X$ such that for all $x\in Y$, $\mathbf{t}(y)=v$. The capacity of vertical category $v$ is $q_{s}^{v}$. We first show that $C_{v}^{h}(Y, q_{s}^{v})\subseteq C_{v}^{h}(Y,1+q_{s}^{v})$. In the computations of $C_{v}^{h}$ with capacity $1+q_{s}^{v}$, the updated capacities of horizontal types at each stage are not lower than the corresponding updated capacities of horizontal types at each stage in the computation of $C_{v}^{h}$ with capacity $q_{s}^{v}$. In addition, the sets of individuals competing for positions in each horizontal type at each stage do not expand when the total capacity increases. Thus, if $x$ is chosen in Step $m$ of $C_{v}^{h}$ when the capacity is $q_{s}^{v}$, then it must be chosen in Step $m$ (or in an earlier step) of $C_{v}^{h}$ when the capacity is $1+q_{s}^{v}$. 
	
We now show that $\mid C_{v}^{h}(Y,1+q_{s}^{v})\mid-\mid C_{v}^{h}(Y,q_{s}^{v})\mid\leq1.$ Recall that $C_{v}^{h}$ is acceptant. There are two cases to consider. 

\paragraph{Case 1} 
Suppose that all $Y=C_{v}^{h}(Y,q_{s}^{v})$. Then, if the total capacity is increased to $1+q_{s}^{v}$, by acceptance, we have $Y=C_{v}^{h}(Y,1+q_{s}^{v})$. Therefore, we have $$\mid C_{v}^{h}(Y,1+q_{s}^{v})\mid-\mid C_{v}^{h}(Y,q_{s}^{v})\mid=0.$$

\paragraph{Case 2}
 Suppose that $C_{v}^{h}(Y,q_{s}^{v})\subset Y$. This implies $\mid C_{v}^{h}(Y,q_{s}^{v})\mid=q_{s}^{v}$. Since all contracts in $Y$ are associated with vertical type $v$, when the capacity increases to $1+q_{s}^{v}$, then we have $\mid C_{v}^{h}(Y,1+q_{s}^{v})\mid=1+q_{s}^{v}$. Hence, we have $$\mid C_{v}^{h}(Y,1+q_{s}^{v})\mid-\mid C_{v}^{h}(Y,q_{s}^{v})\mid=1.$$ 

Since there is no capacity transfer, the monotonicity condition is trivially satisfied. Then, according to the Corollary stated above, $\Phi^{h}$ is stable with respect to $(\mathcal{C}^{h}_{s})_{s\in S}$ and is strategy-proof for individuals. 

Now, we will show that $\Phi^{h}$ eliminates justified envy. Toward a contradiction, suppose that it does not. Then, there must exist a preference profile $P$ such that the result of the COP under $P$, call it $Y$, does not eliminate justified envy. Then, there must exist contracts $x,y\in Y$ such that $(\mathbf{s}(y),\mathbf{t}(y))P_{\mathbf{i}(x)}(\mathbf{s}(x),\mathbf{t}(x))$. Moreover, we have $\mathbf{i}(x) \succ_{\mathbf{s}(y)} \mathbf{i}(y)$ and $\rho(\mathbf{i}(x)) \supseteq \rho(\mathbf{i}(y))$.  Since $(\mathbf{s}(y),\mathbf{t}(y))$ is acceptable to the individual $\mathbf{i}(x)$, she is eligible for the vertical category $\mathbf{t}(y)$.

The individual $\mathbf{i}(x)$ must have offered the contract $(\mathbf{i}(x),\mathbf{s}(y),\mathbf{t}(y))$ in the COP. This contract was rejected by $\mathbf{s}(y)$, while contract $y$ was accepted by $\mathbf{s}(y)$. By Condition (3) in Proposition \ref{technical} in Appendix B, contract $(\mathbf{i}(x),\mathbf{s}(y),\mathbf{t}(y))$ would still be rejected if it were available to institution $\mathbf{s}(y)$. Choosing $y$ while rejecting $(\mathbf{i}(x),\mathbf{s}(y),\mathbf{t}(y))$ contradicts $C^{h}$ being merit-undominated. 

Thus, $\Phi^{h}$ eliminates justified envy.
\end{proof}

\begin{proof}[Proof of Theorem 5]

We prove Theorem 5 by showing that the aggregate choice rule $\mathcal{C}^{hT}$ is in the GS family. In the proof of Theorem 3, we have already shown that $C^{h}$ satisfies the substitutes property and size monotonicity. The responsive choice rule $C_{D}$ trivially satisfies the substitutes property and size monotonicity. Additionally, for $\mathcal{C}^{hT}$, we need to show that the transfer function satisfies the monotonicity condition defined in \cite{aygun2026matching}. The capacity of the "de-reserved" category (D) is simply $q_s^D=r_{OBC}$, where $r_{OBC}$ is the number of unfilled positions in the OBC category preceding category D. This simple transfer function satisfies the monotonicity condition straightforwardly. Thus, $\Phi^{hT}$ is stable with respect to the profile of aggregate choice rule $(\mathcal{C}^{hT})_{s\in S}$ and strategy-proof for individuals. 

Now, we will show that $\Phi^{hT}$ eliminates justified envy. Toward a contradiction, suppose that it does not. Then, there must exist a preference profile $P$ such that the result of the COP under $P$, call it $Y$, does not eliminate justified envy. Then, there must exist contracts $x,y\in Y$ such that $(\mathbf{s}(y),\mathbf{t}(y))P_{\mathbf{i}(x)}(\mathbf{s}(x),\mathbf{t}(x))$. Moreover, we have $\mathbf{i}(x) \succ_{\mathbf{s}(y)} \mathbf{i}(y)$ and $\rho(\mathbf{i}(x)) \supseteq \rho(\mathbf{i}(y))$. 

The individual $\mathbf{i}(x)$ must have offered the contract $(\mathbf{i}(x),\mathbf{s}(y),\mathbf{t}(y))$ in the COP. This contract was rejected by $\mathbf{s}(y)$, while contract $y$ was accepted by $\mathbf{s}(y)$. By Condition (3) in Proposition \ref{technical} in Appendix B, contract $(\mathbf{i}(x),\mathbf{s}(y),\mathbf{t}(y))$ would still be rejected if it were available to institution $\mathbf{s}(y)$. For the aggregate choice rule $\mathcal{C}^{h}_{\mathbf{s}(y)}$ choosing $y$ while rejecting $(\mathbf{i}(x),\mathbf{s}(y),\mathbf{t}(y))$ contradicts the choice rules of categories that are undominated by merit. The responsive choice rule induced by individuals' test scores is also merit-undominated for category-D. Thus,  $\Phi^{hT}$ eliminates justified envy.
\end{proof}

\newpage
\bibliographystyle{aer}

\bibliography{bibliog}

\end{document}